\documentclass[useAMS,usenatbib]{mn2e}

\usepackage{amsmath}
\usepackage{amssymb}
\usepackage{graphicx}
\usepackage{graphics}
\usepackage{float}
\usepackage{epstopdf}
\usepackage{multicol}
\usepackage{hyperref}
\usepackage{breakurl}
\usepackage{textcomp}
\usepackage{lastpage}
\usepackage[flushleft]{threeparttable}
\title[Confirmation of HD 23478's centrifugal magnetosphere]{Confirming HD 23478 as a 
new magnetic B star hosting an H$\alpha$-bright centrifugal magnetosphere\thanks{Based 
on spectropolarimetric observations obtained at the Canada-France-Hawaii Telescope 
(CFHT) which is operated by the National Research Council of Canada, the Institut 
National des Sciences de l'Univers (INSU) of the Centre National de la Recherche 
Scientifique of France, and the University of Hawaii, observations obtained using the 
Narval spectropolarimeter at the Observatoire du Pic du Midi (France), which is operated 
by the INSU, and observations obtained at the Dominion Astrophysical 
Observatory, NRC Herzberg, Programs in Astronomy and Astrophysics, National Research 
Council of Canada.}}
\author[J. Sikora et al.]
	{J.~Sikora,$^{1,2}$ G.~A.~Wade,$^{2}$ D.~A.~Bohlender,$^{3}$ C.~Neiner,$^{4}$ M.~E.~Oksala,$^{4}$
	\newauthor
	M.~Shultz,$^{1,2,10}$ D.~H.~Cohen,$^{5}$ A.~ud-Doula,$^{6}$ J.~Grunhut,$^{7}$ D.~Monin,$^{3}$ S.~Owocki,$^{8}$
	\newauthor
	V.~Petit,$^{9}$ T.~Rivinius,$^{10}$ R.~H.~D.~Townsend$^{11}$\\
$^{1}$Department of Physics, Engineering Physics \& Astronomy, Queen's University, Kingston, ON Canada, K7L 3N6\\
$^{2}$Department of Physics, Royal Military College of Canada, PO Box 17000 Kingston, Ontario K7K 7B4, Canada\\
$^{3}$Herzberg Astronomy and Astrophysics, National Research Council of Canada, 5071 West Saanich Road, Victoria, BC V9E 2E7, \\\hspace{2.5mm}Canada\\
$^{4}$LESIA, Observatoire de Paris, CNRS UMR 8109, UPMC, Universit\'{e} Paris Diderot, 5 place Jules Janssen, 92195 Meudon Cedex, \\\hspace{2.5mm}France\\
$^{5}$Department of Physics \& Astronomy, Swarthmore College, Swarthmore, PA 19081, USA\\
$^{6}$Penn State Worthington Scranton, Dunmore, PA 18512, USA\\
$^{7}$European Organization for Astronomical Research in the Southern Hemisphere, Karl-Schwarzschild-Str. 2, D-85748 Garching bei \\\hspace{2.5mm}M{\"u}nchen, Germany\\
$^{8}$Bartol Research Institute, University of Delaware, Newark, DE 19716, USA\\
$^{9}$Department of Physics \& Space Sciences, Florida Institute of Technology, Melbourne, FL, 32901, USA\\
$^{10}$ESO - European Organisation for Astronomical Research in the Southern Hemisphere, Castilla 19001, Santiago 19, Chile\\
$^{11}$Department of Astronomy, University of Wisconsin-Madison, 2535 Sterling Hall, 475 N Charter Street, Madison, WI 53706, USA}

\begin{document}

\date{6 May 2015 }

\pagerange{000--000} \pubyear{0000}

\maketitle

\label{firstpage}

\begin{abstract}

In this paper we report 23 magnetic field measurements of the B3IV star HD$\,23478$: 12 
obtained from high resolution Stokes $V$ spectra using the ESPaDOnS (CFHT) and Narval 
(TBL) spectropolarimeters, and 11 from medium resolution Stokes $V$ spectra obtained 
with the DimaPol spectropolarimeter (DAO). HD$\,23478$ was one of two rapidly rotating 
stars identified as potential ``centrifugal magnetosphere" hosts based on IR 
observations from the Apache Point Observatory Galactic Evolution Experiment survey. We 
derive basic physical properties of this star including its mass 
($M=6.1^{+0.8}_{-0.7}\,M_\odot$), effective temperature ($T_{\rm eff}=20\pm2\,\text{kK}$), 
radius ($R=2.7^{+1.6}_{-0.9}\,R_\odot$), and age 
($\tau_{\rm age}=3^{+37}_{-1}\,\text{Myr}$). We repeatedly detect weakly-variable 
Zeeman signatures in metal, He and H lines in all our observations corresponding to a 
longitudinal magnetic field of $\langle B_z\rangle\approx-2.0\,\text{kG}$. The rotational 
period is inferred from Hipparcos photometry ($P_{\rm rot}=1.0498(4)\,\text{d}$). Under 
the assumption of the Oblique Rotator Model, our obsevations yield a surface dipole 
magnetic field of strength $B_d\geq9.5\,\text{kG}$ that is approximately aligned with 
the stellar rotation axis. We confirm the presence of strong and broad H$\alpha$ 
emission and gauge the volume of this star's centrifugal magnetosphere to be consistent 
with those of other H$\alpha$ emitting centrifugal magnetosphere stars based on the 
large inferred Alfv\'{e}n to Kepler radius ratio.
\end{abstract}


\section{Introduction}

\begin{table*}
	\caption{Observations of HD$\,23478$ obtained with ESPaDOnS and 
	Narval. The phases are calculated using Eqn. (\ref{eqn:eph}). SNRs per 
	$1.8\,\text{km s}^{-1}$ spectral pixel are reported at $5400\,\text{\AA}$. 
	The three rightmost columns list the longitudinal magnetic field measurements 
	obtained from H$\beta$ and LSD profiles generated from a He+metal line mask 
	along with the detection status according to the criteria of \citet{Donati1997}: 
	definite detection (DD), marginal detection (MD), and no detection (ND) 
	(see Section \ref{mag_field})}.
	\label{obs_tbl}
	\begin{center}
	\begin{tabular*}{1.9\columnwidth}{@{\extracolsep{\fill}}l c c c c c c c}
		\hline
		\hline
		\noalign{\vskip0.5mm}
		HJD & Phase & Total Exp.    & SNR 		    & Instrument & $\langle B_z\rangle_{\rm H\beta}$ & $\langle B_z\rangle_{\rm He+metal}$ & Detection \\
			&		& Time [s]		& [pix$^{-1}$]	&			 & [kG] 						     &  [kG]								  & Status	  \\
		\noalign{\vskip0.5mm}
		\hline
		\noalign{\vskip0.5mm}
		2456884.116 & 0.304 &  560 & 473 & ESPaDOnS & $-1.52\pm0.27$ & $-1.98\pm0.25$ & DD \\
		2456908.146 & 0.194 &  560 & 500 & ESPaDOnS & $-1.68\pm0.24$ & $-2.15\pm0.22$ & DD \\
		2456909.134 & 0.135 &  560 & 381 & ESPaDOnS & $-1.42\pm0.32$ & $-2.33\pm0.29$ & DD \\
		2456909.144 & 0.145 &  560 & 346 & ESPaDOnS & $-1.89\pm0.29$ & $-2.26\pm0.26$ & DD \\
		2456925.512 & 0.736 & 1200 & 402 & Narval   & $-2.19\pm0.37$ & $-1.98\pm0.36$ & MD \\
		2456925.691 & 0.907 & 1200 & 291 & Narval   & $-2.32\pm0.28$ & $-2.49\pm0.30$ & DD \\
		2456962.025 & 0.517 &  560 & 409 & ESPaDOnS & $-0.95\pm0.34$ & $-1.93\pm0.26$ & DD \\
		2456968.937 & 0.101 &  560 & 409 & ESPaDOnS & $-2.13\pm0.31$ & $-2.44\pm0.27$ & DD \\
		2456972.833 & 0.813 &  560 & 353 & ESPaDOnS & $-1.94\pm0.39$ & $-2.52\pm0.37$ & MD \\
		2456972.843 & 0.822 &  560 & 352 & ESPaDOnS & $-2.98\pm0.41$ & $-2.84\pm0.37$ & MD \\
		2456973.099 & 0.066 &  560 & 497 & ESPaDOnS & $-2.05\pm0.27$ & $-2.11\pm0.22$ & DD \\
		2457034.695 & 0.740 &  560 & 461 & ESPaDOnS & $-1.63\pm0.29$ & $-1.99\pm0.27$ & MD \\
		\noalign{\vskip0.5mm}
		\hline \\
	\end{tabular*}
	\end{center}
\end{table*}

Upon the first detection of its magnetic field, the strong, broad, and variable 
H$\alpha$ emission line of the main sequence B2 star $\sigma$ Ori E was suggested to be 
a natural consequence of a  plasma co-rotating with the star well beyond its surface 
\citep{Landstreet1978}. Since then, a significant number ($\gtrsim13$) of magnetic mid- 
to early-B stars (i.e. $T_{\rm eff}<25\,\text{kK}$) exhibiting H$\alpha$ emission were 
discovered \citep{Brown1985,Shore1990a}. These stars are of particular interest because 
they provide a unique means of probing the mass-loss and stellar winds of B-type stars. 
While the strength of the magnetic field likely plays an important role in determining 
if a magnetic B star exhibits H$\alpha$ emission, the key distinction between the stars 
with emission and those without is rotation -- the emission line stars are all rapid 
rotators having rotational periods $\lesssim1.5\,\text{d}$ \citep{Petit2013}.

These correlations between rotation and the presence of H$\alpha$ emission can be 
understood in a general framework in which the magnetospheres of massive stars are 
classified as either dynamic magnetospheres (DM) or centrifugal magnetospheres (CM) 
based on the ability of the field to confine the wind and the degree of criticality of 
the stellar rotation \citep{ud-Doula2002_oth,Petit2013}. \citet{Petit2013}, and more 
recently \citet{Shultz2014}, demonstrate that stars like $\sigma$ Ori E are those with 
the largest magnetospheric volumes: those with the largest Alfv\'{e}n radii combined 
with the smallest Kepler radii. \citet{Shultz2014} estimated that H$\alpha$ emission is 
never observed for typical magnetic B stars with $R_A/R_K\lesssim10$, while it is 
frequently observed in the spectra of stars with ratios above this threshold.

Recently, \citet{Eikenberry2014} identified two emission line B stars from near-infrared 
(nIR) spectra obtained in the context of the Apache Point Observatory Galactic Evolution 
Experiment (APOGEE) survey: HD$\,345439$ and HD$\,23478$. They found that both stars 
exhibited strong and broad emission at all nIR H lines along with relatively 
large projected rotational velocities of $\approx270$ and $125\,\text{km s}^{-1}$, 
respectively. Although lacking any information about their magnetic fields, 
\citet{Eikenberry2014} concluded that the similarities between these stars and $\sigma$ 
Ori E provided reasonable evidence for a magnetospheric origin of the observed nIR 
emission.

The principal aim of the present study is to search for the presence of a 
magnetic field in the brightest of these two stars (HD$\,23478$), and to test 
the hypothesis of \citet{Eikenberry2014} that it is similar to $\sigma$ Ori E in 
that it also belongs to the more general class of H$\alpha$ emitting CM stars. In 
Section \ref{obs} we describe the spectropolarimetric observations used in this study. 
In Section \ref{phys_param} we estimate the physical parameters of HD$\,23478$ using 
photometric data from the literature along with newly-acquired high resolution 
spectroscopy. We then analyze in Section \ref{P_rot} variability in 
Hipparcos epoch photometry in an attempt to infer the star's rotational period. In 
Section \ref{mag_field} we compute the longitudinal magnetic field and search for its 
variability. Section \ref{variability} discusses the spectral line emission and 
variability. In Section \ref{mag} we derive the stellar magnetospheric parameters. 
Finally, in Section \ref{conclusions} we evaluate our hypothesis for a CM origin of 
HD$\,23478$'s emission.

\section{Observations}
\label{obs}
\subsection{High resolution spectropolarimetry}
\label{obs_high}

HD$\,23478$ was observed during eight nights between Aug. 2014 and Jan. 2015 using the 
ESPaDOnS and Narval spectropolarimeters installed at the Canada-France-Hawaii Telescope 
(CFHT) and T\'{e}lescope Bernard Lyot (TBL), respectively. Both instruments have a 
resolving power of $\text{R}\simeq65\,000$ acquiring spectra in circularly polarized 
light spanning the visible range of $3\,600-10\,000\,\text{\AA}$. The Heliocentric 
Julian Dates (HJDs), exposure times, and signal-to-noise ratios (SNRs) of these 
observations are listed in Table \ref{obs_tbl}.

Each circular polarization observation consists of four subexposures which 
the Libre-ESpRIT pipeline \citep{Donati1997} automatically reduces yielding the final 
Stokes $I$ and $V$ spectra. The majority of the observations (10 out of 12) were obtained 
with ESPaDOnS using a total exposure time of $560\,\text{s}$. The remaining two 
observations were obtained with Narval during a single night using a total exposure 
time of 1200 s.

\subsection{Medium resolution H$\beta$ spectropolarimetry}
\label{obs_high}

Eleven spectropolarimetric observations were also obtained using the dimaPol medium 
resolution spectropolarimeter located at the Dominion Astronomical Observatory (DAO) 
from Jan. 2014 to March 2015. The instrument has a resolving power of 
$R\approx10,\,000$ and is optimized to acquire spectra over a wavelength range of 
$4\,700-5\,300\text{\AA}$ \citep{Monin2012}. The HJDs, exposure times, SNRs, and 
longitudinal magnetic field measurements derived from H$\beta$ Stokes $V$ profiles are 
listed in Table \ref{DAO_obs}.

\begin{table}
	\caption{Observations of HD$\,23478$ obtained with dimaPol. The phases are 
	calculated using Eqn. (\ref{eqn:eph}) and the magnetic field measurements correspond 
	to the longitudinal field derived from H$\beta$ Stokes $V$ observations.}
	\label{DAO_obs}
	\begin{center}
	\begin{tabular*}{0.47\textwidth}{@{\extracolsep{\fill}}l c c c c}
		\hline
		\hline
		\noalign{\vskip0.5mm}
		HJD & Phase & Total Exp. & SNR 		   & $\langle B_z\rangle_{\rm H\beta}$\\
			&		& Time [s]	 & [pix$^{-1}$] & [kG] \\
		\noalign{\vskip0.5mm}
		\hline
		\noalign{\vskip0.5mm}
		2456680.700 & 0.537 & 5400 & 300 & $-1.78\pm0.56$\\
		2456681.630 & 0.422 & 3000 & 250 & $-0.92\pm0.72$\\
		2456963.734 & 0.146 & 3000 & 440 & $-1.62\pm0.27$\\
		2456971.888 & 0.911 & 4800 & 400 & $-2.88\pm0.36$\\
		2456972.802 & 0.782 & 4800 & 520 & $-1.82\pm0.26$\\
		2456973.758 & 0.693 & 4800 & 430 & $-1.29\pm0.41$\\
		2456974.846 & 0.731 & 4800 & 410 & $-1.37\pm0.38$\\
		2456976.032 & 0.859 & 4800 & 400 & $-2.08\pm0.29$\\
		2456991.962 & 0.033 & 4800 & 480 & $-1.95\pm0.30$\\
		2456992.711 & 0.748 & 4800 & 310 & $-1.86\pm0.49$\\
		2457085.723 & 0.347 & 4800 & 550 & $-1.11\pm0.33$\\
		\noalign{\vskip0.5mm}
		\hline \\
	\end{tabular*}
	\end{center}
\end{table}

\section{Physical Parameters}
\label{phys_param}
\subsection{SED fitting}
\label{SED_fit}

\begin{figure}
	\centering
	\includegraphics[width=0.99\columnwidth]{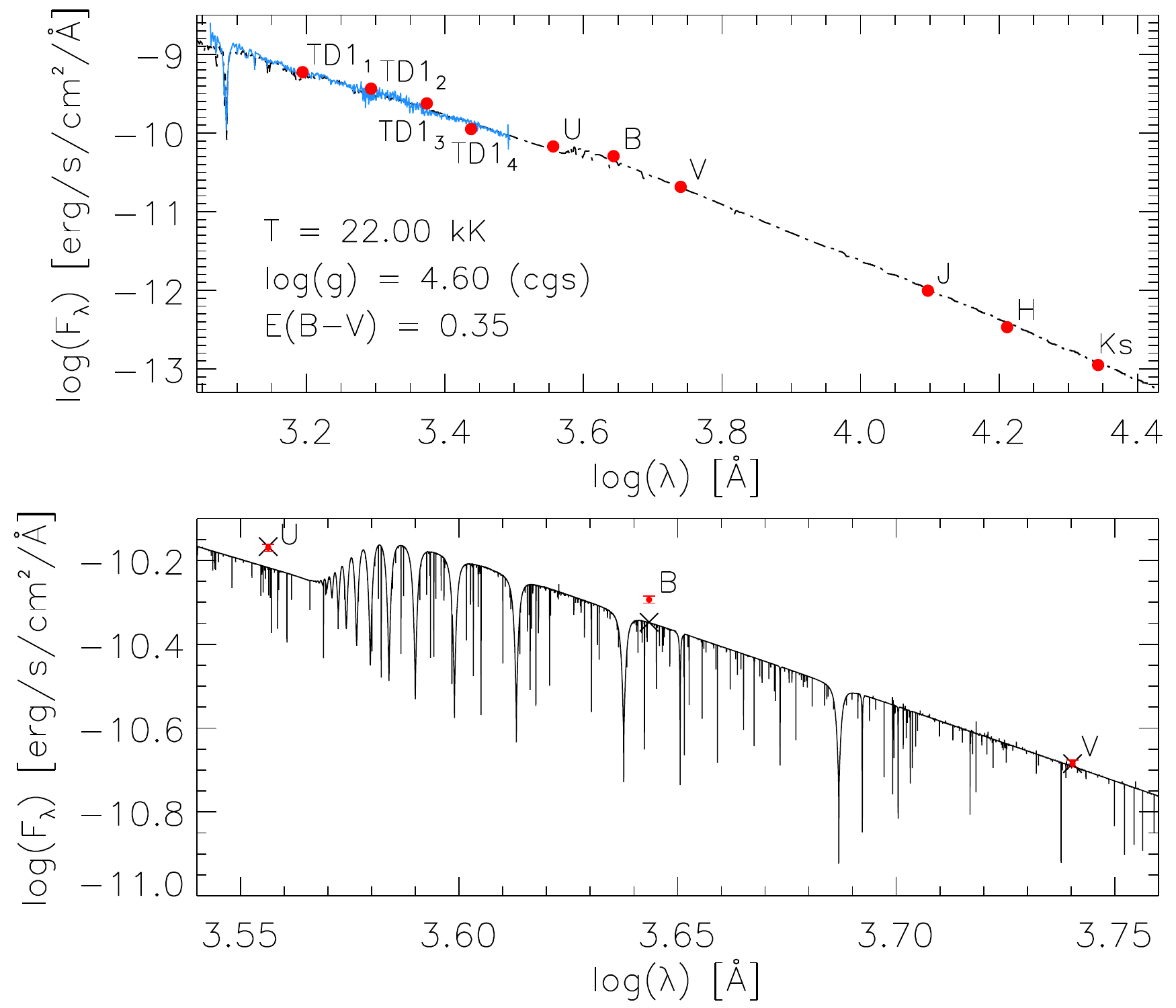}
	\caption{\emph{Top:} Comparing the IUE spectra (solid blue) and monochromatic 
	fluxes (red points) with the best-fitting TLUSTY model (dashed black) convolved by a 
	Gaussian profile to match the rebinned IUE resolution. \emph{Bottom:} Theoretical 
	$U$, $B$, and $V$ fluxes (black crosses) were calculated by convolving 
	the unbroadened TLUSTY models (solid black) with the filter's associated 
	transmission functions.}
	\label{SED}
\end{figure}

\begin{figure}
	\centering
	\includegraphics[width=0.99\columnwidth]{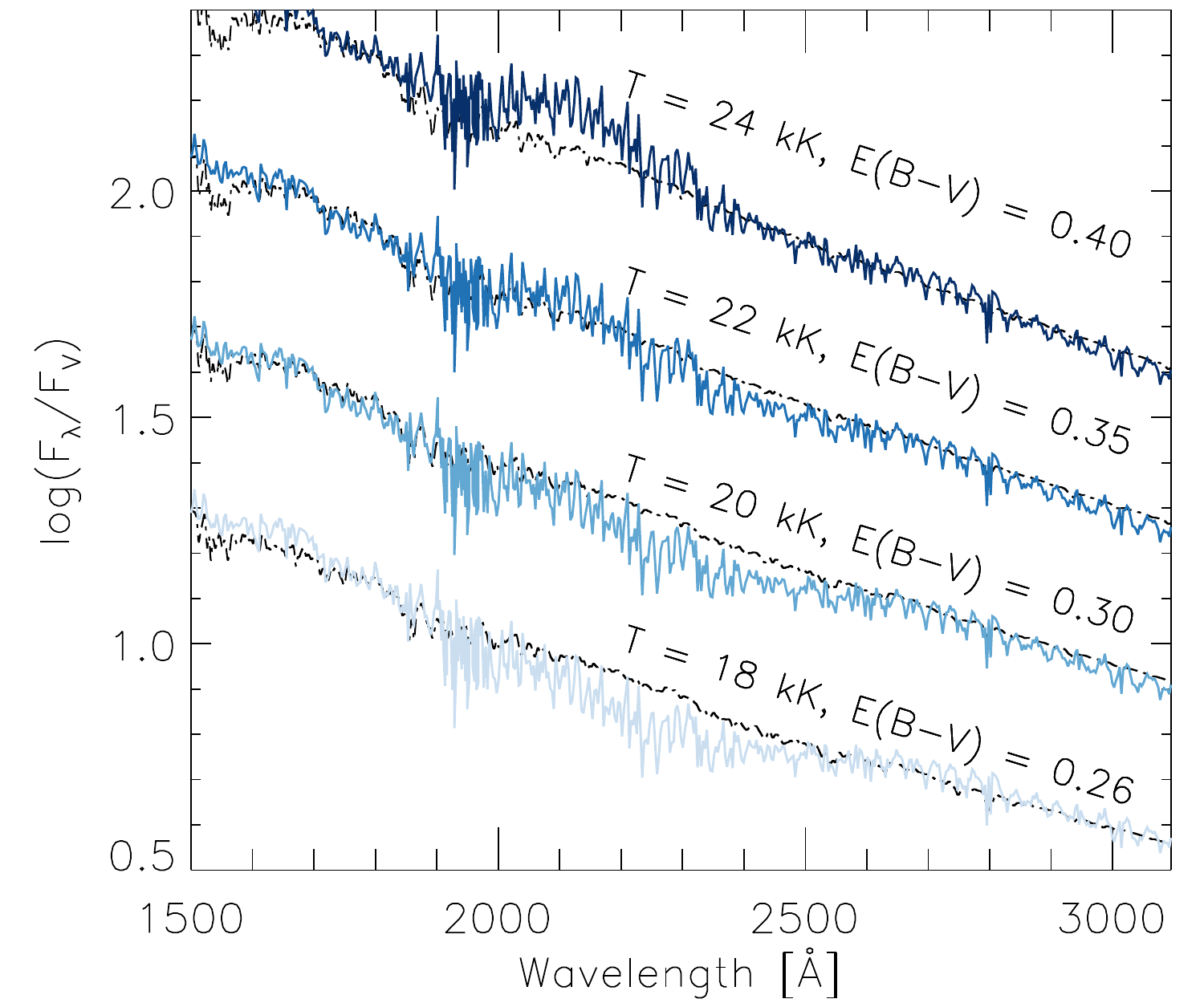}
	\caption{Comparisons between the IUE LWP spectrum (solid blue) and model SEDs 
	(dashed black) at various effective temperatures and colour excesses. Both the 
	models and the observations are normalized to the $V$ flux ($F_V$).}
	\label{IUE_SED}
\end{figure}

Photometry spanning a wide range of wavelengths is available for HD$\,23478$ 
thereby providing a means of constraining both the effective temperature and surface 
gravity. These observations include monochromatic Johnson $UBV$ 
\citep{Anderson2012}, 2MASS $JHKs$ \citep{Cutri2003}, and $TD1$ ultraviolet 
\citep{Thompsons1978} photometric measurements. Short (SWP imager: 
$1150-1975\,$\AA) and long (LWP imager: $1900-3125\,$\AA) wavelength ultraviolet 
broadband measurements were also recorded by the International Ultraviolet Explorer 
(IUE). The high resolution ($\delta\lambda=0.1\,$\AA) IUE data were rebinned at 
$\delta\lambda=3\,\text{\AA}$ using the prescription of \citet{Solano1998} in order to 
increase the SNR per pixel. All of the observed magnitudes were converted to physical 
fluxes using published zero points for Johnson and 2MASS filters 
\citep{Bessell1979,Cohen2003}; both $TD1$ and IUE measurements were obtained in the 
appropriate units of $\text{ergs}\,\text{s}^{-1}\,\text{cm}^{-2}\,\text{\AA}^{-1}$. 
Dereddening was applied using the method of \citet{Cardelli1989} using a typical 
total-to-selective extinction of $R(V)\equiv A(V)/E(B-V)=3.1$.

The photometric observations were compared with the non-local thermodynamic equilibrium 
(NLTE) TLUSTY BSTAR2006 synthetic energy distribution (SED) grid \citep{Lanz2007} for 
effective temperatures $T_{\rm eff}$ ranging from $15-30\,\text{kK}$ and surface 
gravities $\log{g}$ (cgs) from $3-4.75$. The published grid has $T_{\rm eff}$ and 
$\log{g}$ resolutions of $1000\,\text{K}$ and $0.25$, respectively, that were 
linearly interpolated to a finer grid of $250\,\text{K}$ and $0.05$ resolution. 
Solar abundances and a $2\,\text{km s}^{-1}$ turbulence velocity were assumed for 
the model grid. The NLTE model SEDs have a resolution $\gtrsim0.002\,\text{\AA}$ -- 
much higher than the rebinned IUE observations. We reduced the resolution of the model 
spectra by convolving with a Gaussian profile assuming a resolving power of 
$R=(1150\,\text{\AA})/(3\,\text{\AA})\approx380$ to match the $3\,\text{\AA}$ IUE 
bin-width at $\lambda=1150\,\text{\AA}$. Theoretical monochromatic fluxes were calculated 
by convolving Johnson $UBV$ and 2MASS $JHKs$ transmission functions 
\citep{Landolt2007,Cohen2003} with the unbroadened model spectra.

The best fitting NLTE model, shown in Fig. \ref{SED}, was found to have an 
effective temperature, surface gravity, and colour excess of 
$T_{\rm eff}=22.00\,\text{kK}$, $\log{g}=4.60\,\text{(cgs)}$, and $E(B-V)=0.35$ 
yielding a reduced $\chi^2$ value of $1.70$. The monochromatic fluxes were found to be 
adequately reproduced by a wide range of $T_{\rm eff}$, $\log{g}$, and $E(B-V)$ values. 
Changes in the surface gravity of each model SED were found to primarily 
affect the depths of the absorption lines, resulting in minimal changes in the 
calculated monochromatic fluxes. The IUE spectra provided the greatest constraint on 
the fitting parameters. In particular, we found relatively large discrepancies between 
the IUE LWP spectra and the model SEDs having $T_{\rm eff}\lesssim20\,\text{kK}$ and 
$T_{\rm eff}\gtrsim24\,\text{kK}$ at $2,000\,\lesssim\lambda\lesssim2,500\,\text{\AA}$. 
This is clearly a consequence of the respectively lower and higher values of the 
required extinction associated with these models. In Fig. \ref{IUE_SED}, we show the LWP 
spectrum dereddened in order to fit models of 18, 20, 22, and 24 kK. The 
$2,000-2,500\,\text{\AA}$ region shows growing discrepancies for the lower and higher 
temperatures. This leads us to prefer $T_{\rm eff}=22\,\text{kK}$ based on the SED 
fitting.

\begin{table}
	\caption{Stellar parameters of HD$\,23478$.}
	\label{param_tbl}
	\begin{center}
	\begin{tabular*}{0.35\textwidth}{@{\extracolsep{\fill}}l r}
		\hline
		\hline
		\noalign{\vskip1mm}
		Sp. Type$^{1}$ & B3IV\\
		$\pi$ [mas]$^{2}$ & $4.99\pm0.62$\\
		$d$ [pc] & $200^{+29}_{-22}$\vspace{2mm}\\
		Photometric &\\
		\hline
		\noalign{\vskip1mm}
		$V$ [mag]$^{3,4}$ & $6.67^{+0.01}_{-0.07}$\vspace{0.8mm}\\
		$E(B-V)$ [mag] & $0.30_{-0.04}^{+0.05}$\vspace{0.8mm}\\
		$BC$ [mag] & $-1.97^{+0.24}_{-0.22}$\vspace{0.8mm}\\
		$M_V$ [mag] & $-0.77^{+0.26}_{-0.36}$\vspace{0.8mm}\\
		$M_{\rm bol}$ [mag] & $-2.74^{+0.51}_{-0.58}$\vspace{2mm} \\
		Physical &\\
		\hline
		\noalign{\vskip1mm}
		$T_{\rm eff}\,[\text{kK}]$ & $20\pm2$\vspace{0.8mm}\\
		$\log(g)\,[cgs]$ & $4.2\pm0.2$\vspace{0.8mm}\\
		$v\sin{i}\,[\text{km s}^{-1}]$ & $140\pm10$\vspace{0.8mm}\\
		$\log{L/L_\odot}$ & $3.0\pm0.2$\vspace{0.8mm}\\
		$M/M_\odot$ & $6.1^{+0.8}_{-0.7}$\vspace{0.8mm}\\
		$R/R_\odot$ & $2.7_{-0.9}^{+1.6}$\vspace{0.8mm}\\
		$\tau_{\rm age}\,[\text{Myr}]$ & $3^{+37}_{-1}$\vspace{0.8mm}\\
		$P_{\rm rot}\,[\text{d}]$ & $1.0498\pm0.0004$\vspace{0.8mm}\\
		$v_{r}\,[\text{km s}^{-1}]$ & $17\pm5$\vspace{2mm}\\
		\hline
	\end{tabular*}\par
	\begin{tablenotes}
	\small
	\item Table references: $^1$\citet{Blaauw1963}, $^2$\citet{VanLeeuwen2007}, 
	$^3$\citet{Syfert1960}, $^4$\citet{Harris1956_oth}.
	\end{tablenotes}
	\end{center}
\end{table}

\begin{figure}
	\centering
	\includegraphics[width=0.99\columnwidth]{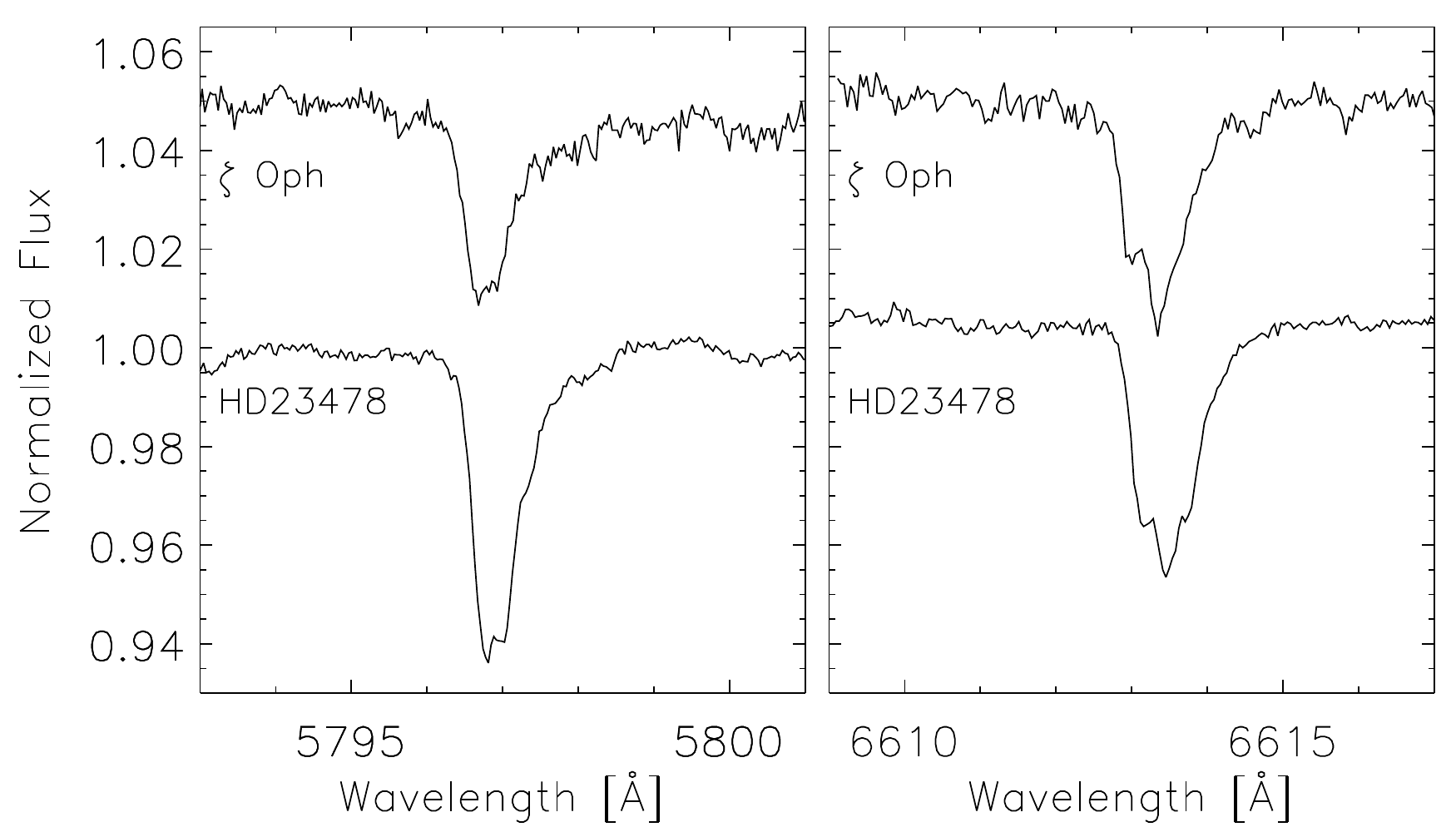}
	\caption{Comparisons between two DIBs found in the spectra of HD$\,23478$ and 
	$\zeta\,$Oph.}
	\label{DIBs}
\end{figure}

\begin{figure*}
	\centering
	\includegraphics[width=2.1\columnwidth]{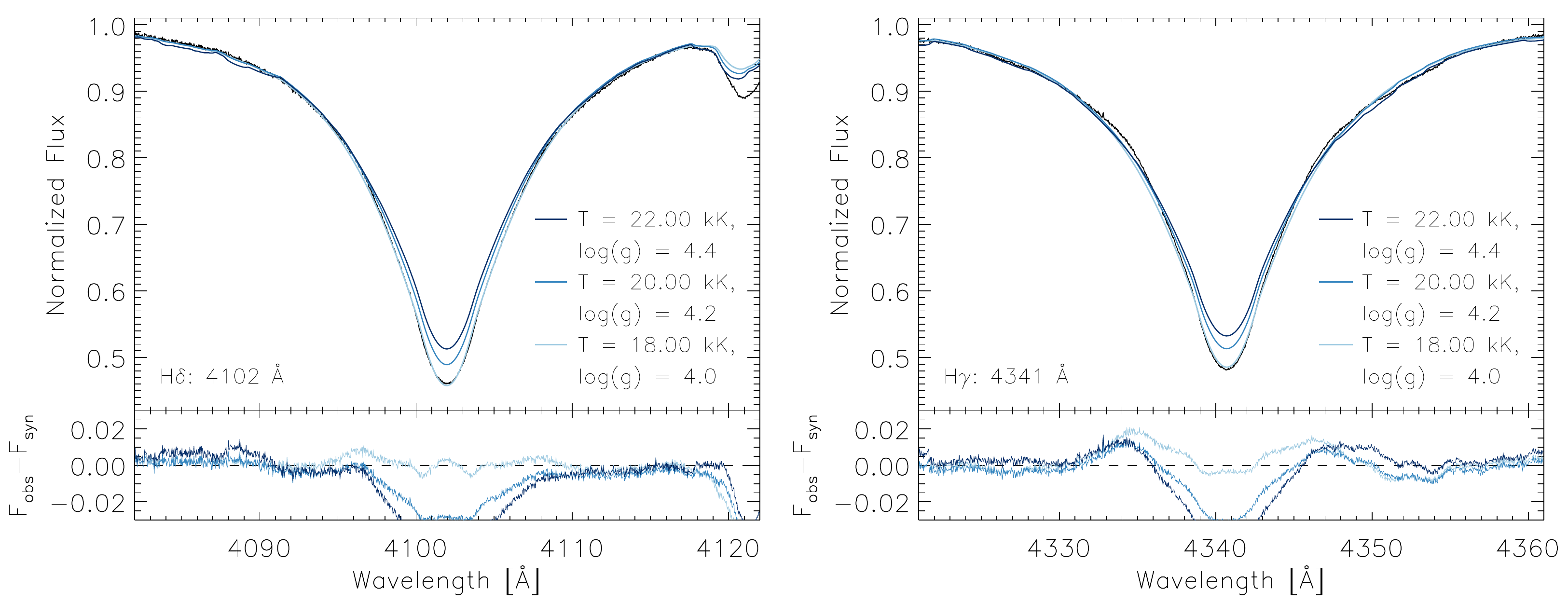}
	\caption{Comparisons between TLUSTY models (blue) and the averaged observed spectrum 
	(black) for H$\delta$ (left) and H$\gamma$ (right). The bottom frames show the 
	residuals between the observations and the synthetic spectra 
	($F_{\rm obs}-F_{\rm syn}$)}
	\label{spec_line}
\end{figure*}

\begin{figure*}
	\centering
	\includegraphics[width=2.1\columnwidth]{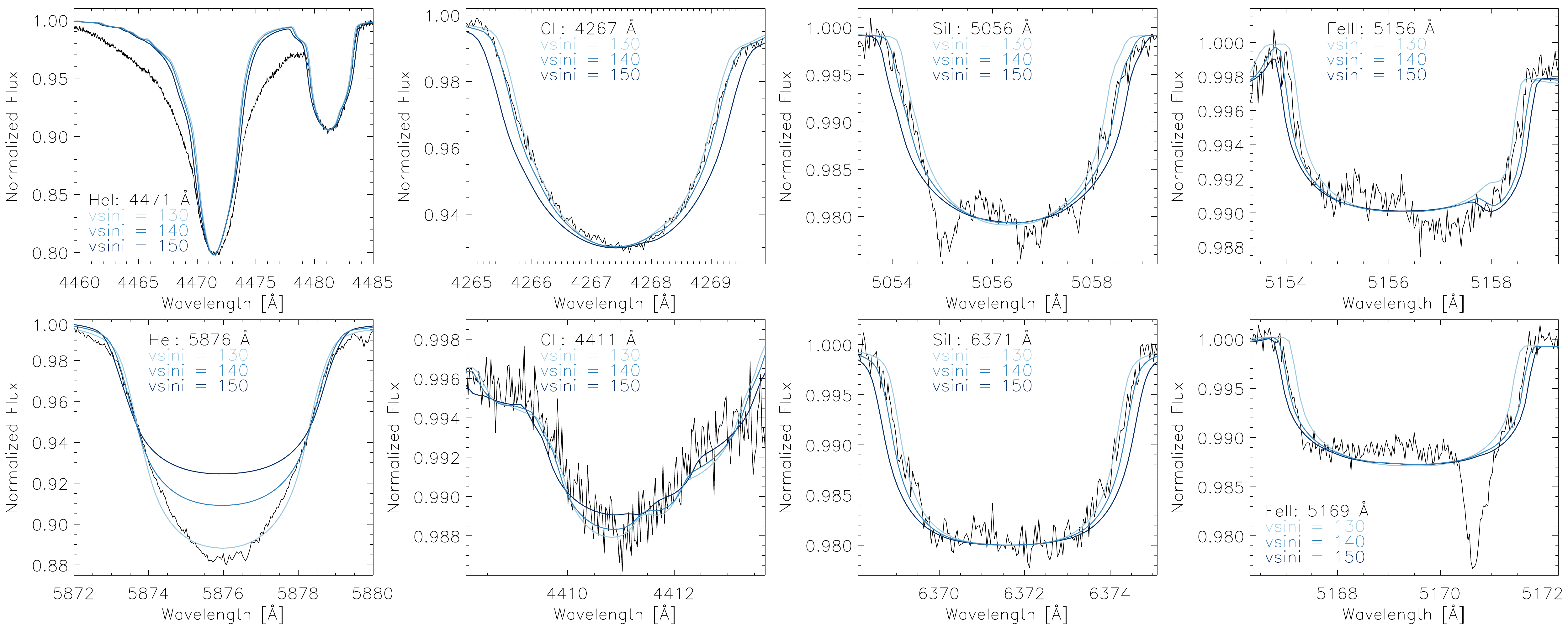}
	\caption{$T_{\rm eff}=22\,\text{kK}$ and $\log{g}=4.4$ TLUSTY models with a range of 
	$v\sin{i}$ values compared with the averaged observed spectrum (black). The 
	abundances of each element were adjusted in order to yield the best overall fit to 
	the synthetic spectra. DIBs in both the Si~{\sc ii} $\lambda5056$ and 
	Fe~{\sc ii} $\lambda5169$ lines are visible at $5055\,\text{\AA}$ and 
	$5170.5\,\text{\AA}$.}
	\label{brd_srch}
\end{figure*}

\subsection{Spectral line fitting}
\label{line_fit}

Spectral line fitting can provide a further constraint on $T_{\rm eff}$ and $\log{g}$ 
through comparisons between the obtained ESPaDOnS and Narval spectra and synthetic model 
grids. We found that HD$\,23478$'s metallic line spectrum was not suitable for tuning the 
atmospheric parameters due to a number of characteristics. The strongest metal lines have 
shallow depths of less than $10\%$ of the continuum with only 3 lines deeper than $5\%$. 
This difficulty was alleviated to some extent by using averaged spectra thereby 
increasing the SNR by a factor of $\approx3.5$.

We also noticed that many line profiles were distorted. Sharp asymmetric 
features having depths $\lesssim10\,\%$ of the continuum are prevalent in many spectral 
orders (Fig. \ref{DIBs}). These were identified as diffuse interstellar bands (DIBs) by 
conferring with \citet{Hobbs2008,Hobbs2009} and by directly comparing the spectra of 
HD$\,23478$ and $\zeta\,$Oph, a star which is well known to exhibit strong DIB 
features \citep[e.g.][]{Walker2000_oth}.

In addition to the DIB features, many metal lines show peculiar strengths and 
distorted profiles, likely due to non-solar surface abundances and nonuniform 
distributions of the abundances of these elements. As a consequence, modelling the 
metallic spectrum to constrain the atmospheric parameters would require a detailed 
analysis which is outside the scope of this work. Therefore we opted to model 
the Balmer lines to provide further constraint on $T_{\rm eff}$ and $\log{g}$ because 
they are more weakly influenced by abundance non-uniformities. Specifically, 
H$\gamma$ and H$\delta$ were chosen because of their relative insensitivity to 
emission (as compared to H$\alpha$) and because they are not directly 
adjacent to other strong absorption features thus reducing normalization errors.

We relied primarily upon spectral models calculated using SYNPLOT, the IDL 
wrapper for SYNSPEC \citep{Hubeny2011}, along with NLTE TLUSTY model atmospheres 
\citep{Lanz2007}. The grid was linearly interpolated such that a finer resolution 
of $\delta T_{\rm eff}=250\,\text{K}$ and $\delta\log{g}=0.1\,\text{(cgs)}$ was 
obtained. The synthetic spectra were then convolved with a Gaussian profile associated 
with the characteristic $R=65,000$ resolving power of the ESPaDOnS and Narval 
spectropolarimeters, corresponding to a velocity resolution of 
$\delta v\approx4.6\,\text{km s}^{-1}$. It is noted that changes in the He abundance can 
affect the wings of the Balmer lines \citep{Leone1997}. When calculating the NLTE H line 
profiles, we use the minimum He abundance of $[{\rm He}/{\rm H}]=-0.58$ required to fit 
a sample of averaged He~{\sc i} lines 
\citep[where $\lbrack{\rm He}/{\rm H}\rbrack_\odot=-1.01$;][]{Grevesse1996}.

As shown in Fig. \ref{spec_line}, the best effective temperature according to 
the SED fitting ($22\,\text{kK}$) reproduces the Balmer line wings reasonably well with 
$\log{g}=4.4$. However, the core depths are seriously underestimated by the model. By 
reducing $T_{\rm eff}$, good agreement between the observed and computed wings and core 
is achieved for $T_{\rm eff}=18\,\text{kK}$ and $\log{g}=4.0$.

We therefore find that the Balmer lines and SED yield different solutions to the 
atmospheric parameters. Based on our analysis, we conclude that the effective 
temperature of HD$\,23478$ is likely in the range $18-22\,\text{kK}$ and its surface 
gravity between $4.0$ and $4.5$. Further refinement of these parameters will require a 
more detailed future analysis.

The colour excess of $E(B-V)=0.30^{+0.05}_{-0.04}$ inferred for the 
$T_{\rm eff}=20\pm2\,\text{kK}$ range is in agreement with both the $0.28$ and 
$0.34$ values reported (without consideration of uncertainties) by 
\citet{Voshchinnikov2012} and \citet{Beeckmans1980}. For $T_{\rm eff}=22\,\text{kK}$ 
however, a slightly higher $E(B-V)=0.35$ is required. As was expected, $T_{\rm eff}$ 
was indeed found to be closely coupled with the colour excess.

\subsubsection{Rotational broadening}
\label{rot_brd}

To estimate the rotational broadening, we compared observed phase-averaged He 
and metal profiles with synthetic profiles computed assuming $T_{\rm eff}=22\,\text{kK}$. 
We adjusted the projected rotational velocity, $v\sin{i}$, as well as the abundance of 
each element in order to obtain a reasonable fit to the line profiles. Generally, a 
$v\sin{i}$ of $140\pm10\,\text{km s}^{-1}$ yielded the best overall fit to the line 
profiles \citep[consistent with that of][]{Eikenberry2014}. Fits to selected lines for 
several $v\sin{i}$ values are shown in Fig. \ref{brd_srch}. The inclusion of either 
macroturbulent or microturbulent broadening did not significantly improve the fits. The 
distorted shapes of many line profiles are apparent. We also note the remarkably poor 
fit to the He~{\sc i} $\lambda4471$ line. This discrepancy cannot be relieved by changing 
$T_{\rm eff}$ or $\log{g}$. It is reproduced in a number of other He~{\sc i} lines. We 
suspect that it is a consequence of stratification of He in the atmosphere of this 
star \citep[e.g.][]{Bohlender1989,Farthmann1994}.

\subsection{Hertzsprung-Russell Diagram}

The mass and age of HD$\,23478$ were estimated by comparing its location on the 
Hertzsprung Russell diagram (HRD) with theoretical isochrones and evolutionary tracks.

The luminosity of HD$\,23478$ required to determine its location on the HRD was 
calculated using the de-reddened visual magnitude of 
$V=5.62^{+0.01}_{-0.07}\,\text{mag}$ where the color excess of $E(B-V)=0.30$ discussed 
in Section \ref{line_fit} was used. The calculation of the absolute visual magnitude, 
$M_V$, required a bolometric correction which was approximated by the temperature and 
surface gravity relation by \citet{Nieva2013} applicable to stars with 
$15.8\leq T_{\rm eff}\leq34.0\,\text{kK}$. The luminosity was then found to be 
$L=1.0_{-0.4}^{+0.7}\times10^3\,L_{\odot}$ where the primary contributions to the 
uncertainty are from the temperature (through the $BC$ calibration) and parallax (i.e. 
the distance). The temperature of $T_{\rm eff}=20\pm2\,\text{kK}$ yielded by the 
SED fitting was used in the final HRD placement.

The comparisons used Geneva model grids of isochrones ranging in stellar age from 
$3.16\,\text{Myr}$ to $12.9\,\text{Gyr}$ generated by \citet{Ekstrom2012}. The associated 
grid of evolution tracks contains models ranging in mass from 0.8 to $120\,M_{\odot}$; 
both the isochrone and evolution grids assume a solar metallicity of $Z=0.014$. An 
identical resolution, $Z=Z_\odot$ grid which included rotational effects for a star 
having $v/v_{\rm crit}=0.4$ \citep{Georgy2013} was also used in the analysis. However, 
the comparisons between the derived mass and radius of HD$\,23478$ using both the 
rotating and non-rotating model grids yielded negligible differences of less than 5\%.

\begin{figure}
	\centering
	\includegraphics[width=0.999\columnwidth]{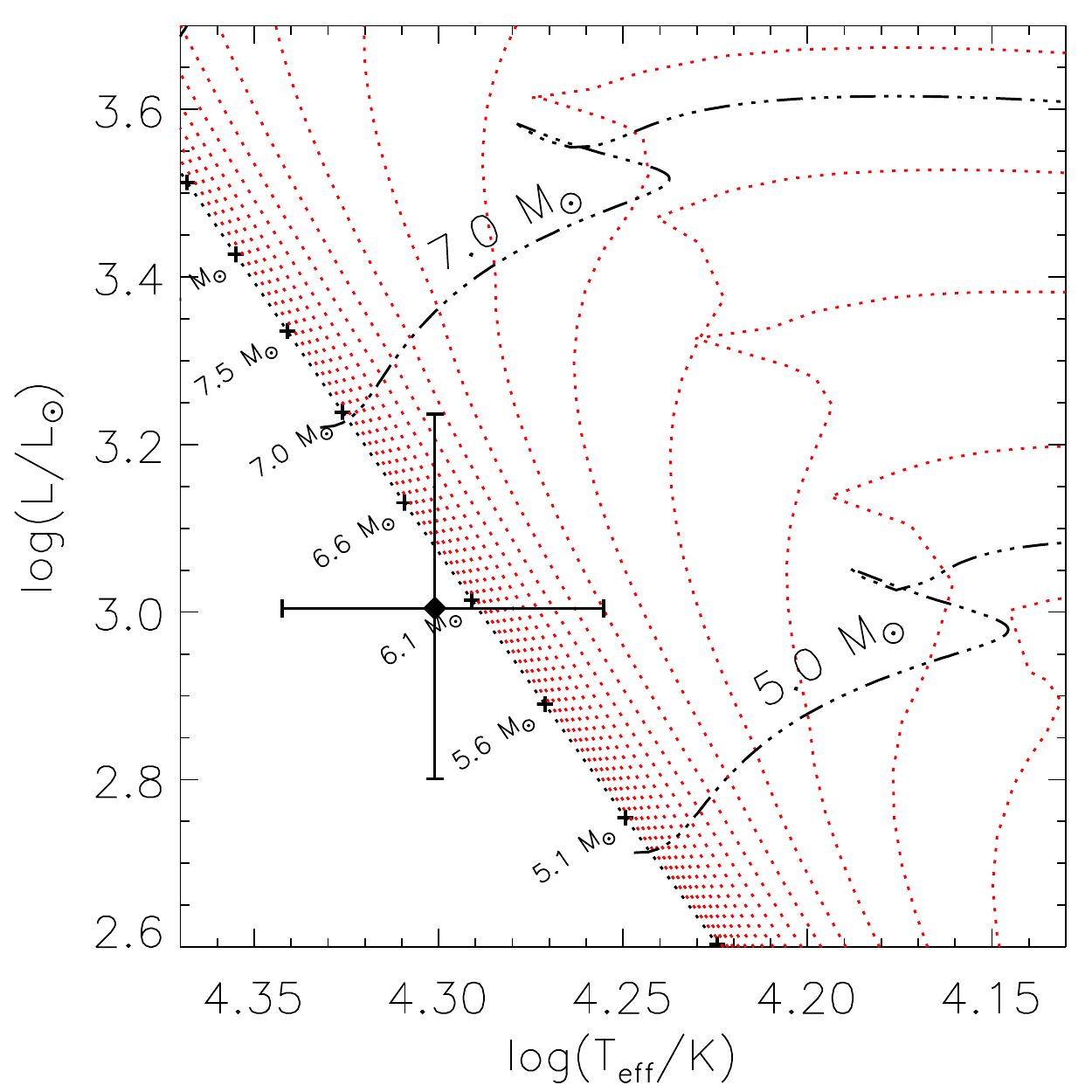}
	\caption{The position of HD$\,23478$ is indicated by the black diamond. The 
	evolutionary tracks (black dotted-dashed lines) and isochrones (red dotted lines) 
	assume a non-rotating star of solar metallicity \citep{Ekstrom2012}.}
	\label{hrd}
\end{figure}

The position of HD$\,23478$ on the HRD is shown in Fig. \ref{hrd} where the 
nearby stellar evolution tracks are shown as dotted black lines and the main sequence 
masses are indicated by black `plus' symbols. The isochrones (indicated by red dotted 
lines), $\log{L}$, and $\log{T_{\rm eff}}$ imply a range in the estimated age and radius 
for the star of $\tau=3^{+37}_{-1}\,\text{Myr}$ and 
$R=2.8^{+1.3}_{-0.2},R_\odot$, respectively. Likewise, the main sequence mass of 
HD$\,23478$ is estimated to be $M=6.1^{+0.8}_{-0.7}\,M_\odot$. A similar radius of 
$R=2.7^{+1.6}_{-0.9},R_\odot$ is derived from $L$ and $T_{\rm eff}$ using the 
Stefan-Boltzmann law; we adopt this value because of its larger uncertainty. We note 
that these mass and radius estimates imply a surface gravity of 
$\log{g}=4.3^{+0.5}_{-0.4}$ which is consistent with the value derived from the 
spectral line fitting procedure described in Section \ref{line_fit}.

\section{Rotational period}
\label{P_rot}
The Hipparcos Epoch Photometry catalogue \citep{Perryman1997} contains 75 measurements 
of HD$\,23478$ spanning a time period of three years with minimum and maximum time 
separations of approximately 20 minutes and 174 days, respectively. The measurements 
are therefore sufficient for detecting photometric variability across a wide range of 
time scales associated with, for example, stellar rotation. All of the uncertainties 
are reported to be between $0.005$ and $0.011$ magnitudes with a mean value of $0.007$. 
Out of the 75 observations, three report the presence of anomalous measurements 
between the FAST and NDAC Data Reduction Consortia. Furthermore, one observation 
reports a magnitude that is $2.7\langle\sigma_{H_p}\rangle$ larger than 
$\langle H_p\rangle$ and is likely an outlier. The exclusion of any of the flagged or 
outlying data points had an insignificant effect on the estimated rotational period with 
a maximal deviation of $\Delta P<5\,\text{s}$.

\begin{figure}
	\centering
	\includegraphics[width=0.99\columnwidth]{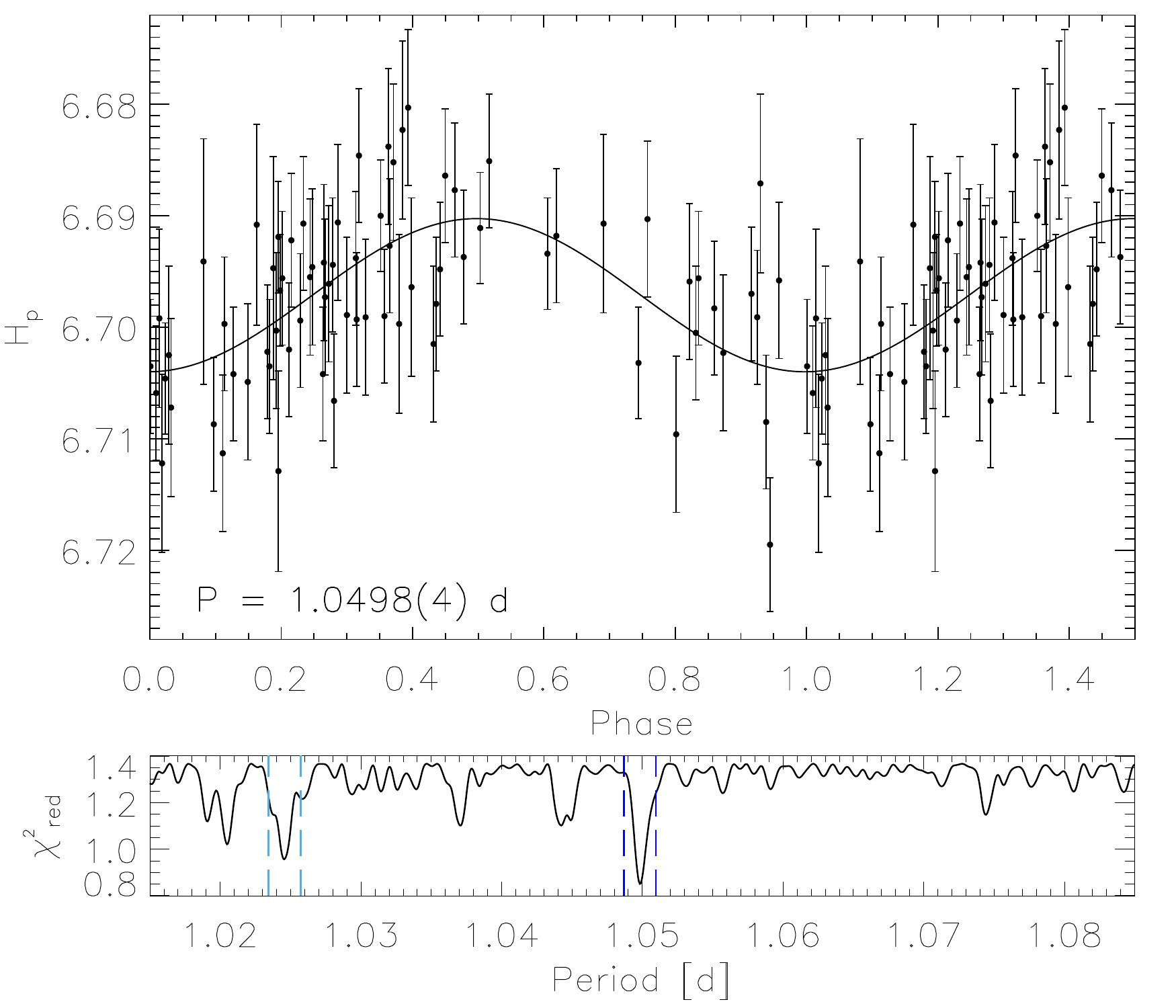}
	\caption{\emph{Top:} Hipparcos Epoch Photometry phased with a period of 
	$1.0498\text{d}$. \emph{Bottom:} A subsample of the full reduced $\chi^2$ 
	versus period centered on the period of the best fitting sinusoid. The 
	$1.0498\,\text{d}$ minimal-$\chi^2$ solution is bounded by the dashed dark blue 
	lines. The dashed light blue lines ($P_{\rm rot}=1.02454\,\text{d}$) indicate the 
	only other $\chi^2$ peak falling below our chosen $3\sigma$ threshold.}
	\label{Hipp}
\end{figure}

The period of the Epoch Photometry observations was found by assuming a sinusoidal 
fit and calculating the reduced $\chi^2$ distribution for periods ranging from 
0 to 5 days with a resolution of $\delta P\sim5\,\text{sec}$. This yielded a range of 
minimal-$\chi^2$ solutions with the global minimum having a reduced $\chi^2$ value of 
$\chi_{\rm red}^2=0.85$, occuring at a period of $P_{\rm rot}=1.0498\,\text{d}$. The 
Hipparcos measurements phased at this best-fitting period are shown in Fig. \ref{Hipp} 
along with a subsample of the full periodogram centered on 
$P_{\rm rot}=1.0498\,\text{d}$.

Assuming a $3\sigma$ confidence limit corresponding to an uncertainty of 
$\delta P=0.0004\,\text{d}$, we find that only one other period has a $\chi^2$ value 
below this threshold at $P=1.02454\,\text{d}$ where the $\chi^2$ peak is shown 
in Fig. \ref{Hipp} (bottom) bounded by the dashed dark blue lines. The 
$1.02454\,\text{d}$ period does not result in a better fit compared with the 
$1.0498\,\text{d}$ period; moreover, a similar analysis has previously been applied to 
ground-based photometry resulting in a photometric period of $1.0499\,\text{d}$ (for 
which no uncertainty is reported) \citep{Jerzykiewicz1993}. We therefore adopt the 
$1.0498\,\text{d}$ period along with the ephemeris 
\begin{equation}
\label{eqn:eph}
JD=2448700.606\pm1.0498(4)\cdot E
\end{equation}
where the reference JD ($2448700.606$) corresponds to the epoch of $H_p$ maximum.

When phased with this ephemeris, the $H_p$ data exhibit a roughly sinusoidal variation 
with a full amplitude of $0.015\,\text{mag}$ and extrema located at phases of 
$0.00\pm0.05$ and $0.50\pm0.05$.

\begin{figure}
	\centering
	\includegraphics[width=0.99\columnwidth]{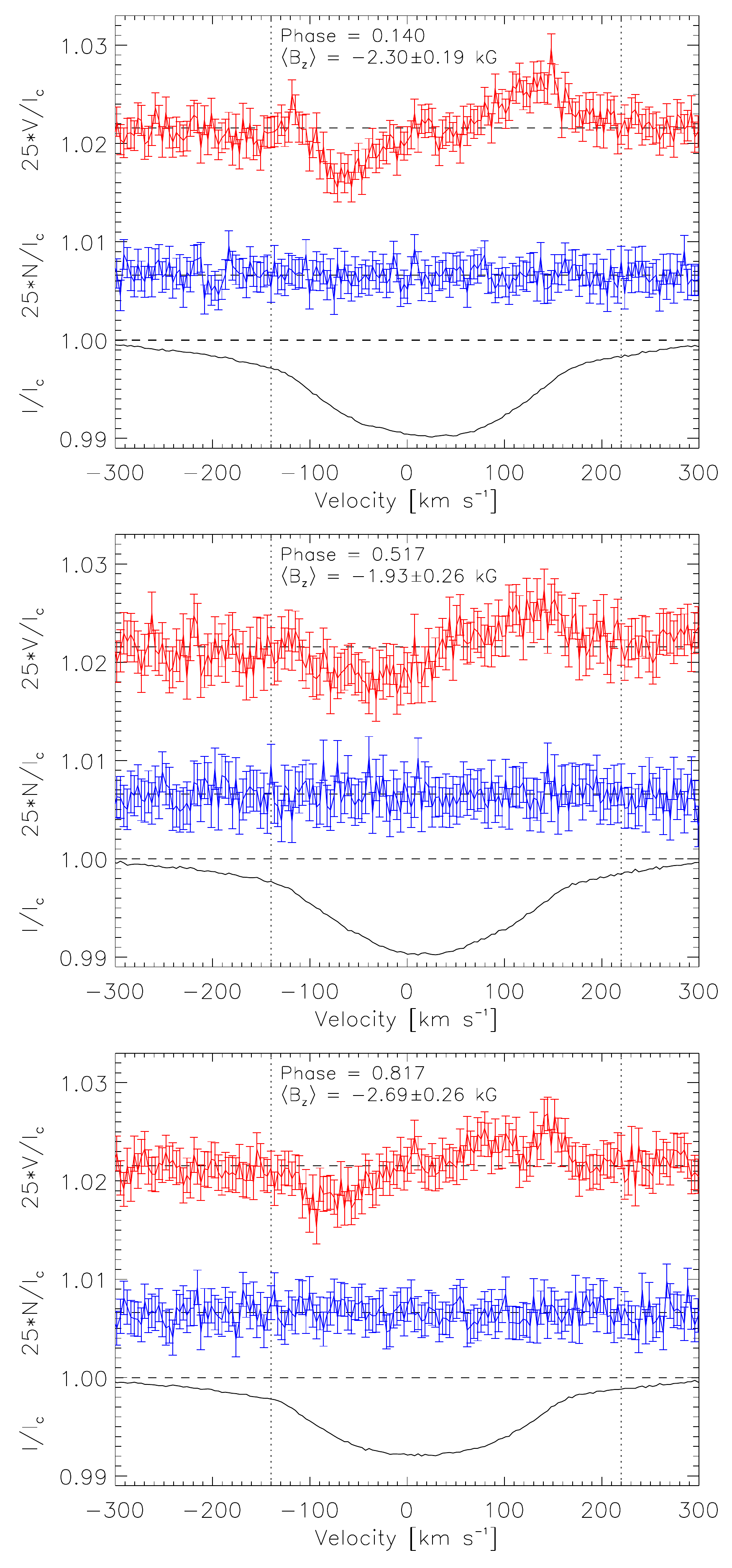}
	\caption{Examples of He+metal LSD profiles at 3 phases showing Zeeman 
	signatures in Stokes $V$ (top, red), along with associated Stokes $I$ (bottom, 
	black) and diagnostic null (middle, blue) profiles. Vertical dotted lines indicate 
	the integration range.}
	\label{LSD}
\end{figure}

As discussed by \citet{Jerzykiewicz1993}, several possibilities exist for the origin of 
the $1.0498\,\text{d}$ photometric period. Slowly pulsating B stars typically have 
periods $\sim0.5-5.0\,\text{d}$ and can be conclusively identified by the presence of 
multiperiodicity \citep{Waelkens1991}. \citet{Jerzykiewicz1993} noted that 
$g$-mode pulsations could potentially explain the observed variability. Another 
explanation is that HD$\,23478$ is an eclipsing binary with a dimmer, less-massive 
companion for which no spectral signature is found. \citet{Jerzykiewicz1993} rejected 
this possibility arguing that radial velocity variations $\gtrsim80\,\text{km s}^{-1}$ 
would be exhibited in this scenario. The observations analyzed here do not exhibit any 
variations $\geq5\,\text{km s}^{-1}$.

Assuming that the photometric variations are instead caused by rotational modulation, 
the inclination angle of the stellar rotation axis, $i$, can be inferred from 
the projected rotational velocity, $v\sin{i}=140\pm10\,\text{km s}^{-1}$, and the 
stellar radius, $R=2.7^{+1.6}_{-0.9}\,R_\odot$, resulting in an inclination of 
$i=69^{+21\,\circ}_{-10}$ where the uncertainty is propagated from the $3\sigma$ value of 
$\delta\,P_{\rm rot}=0.0004\,\text{d}$ and estimated errors in $R$ and $v\sin{i}$. These 
values imply a stellar rotational velocity of $v=150_{-20}^{+30}\,\text{km s}^{-1}$ which 
is consistent with typical values for B stars \citep{McNally1965}. Therefore, in 
agreement with \citet{Jerzykiewicz1993}, we conclude that the observed Hipparcos 
photometric variations are mostly likely generated by the star's rotation.

\section{Magnetic Field}
\label{mag_field}

\begin{figure}
	\centering
	\includegraphics[width=0.99\columnwidth]{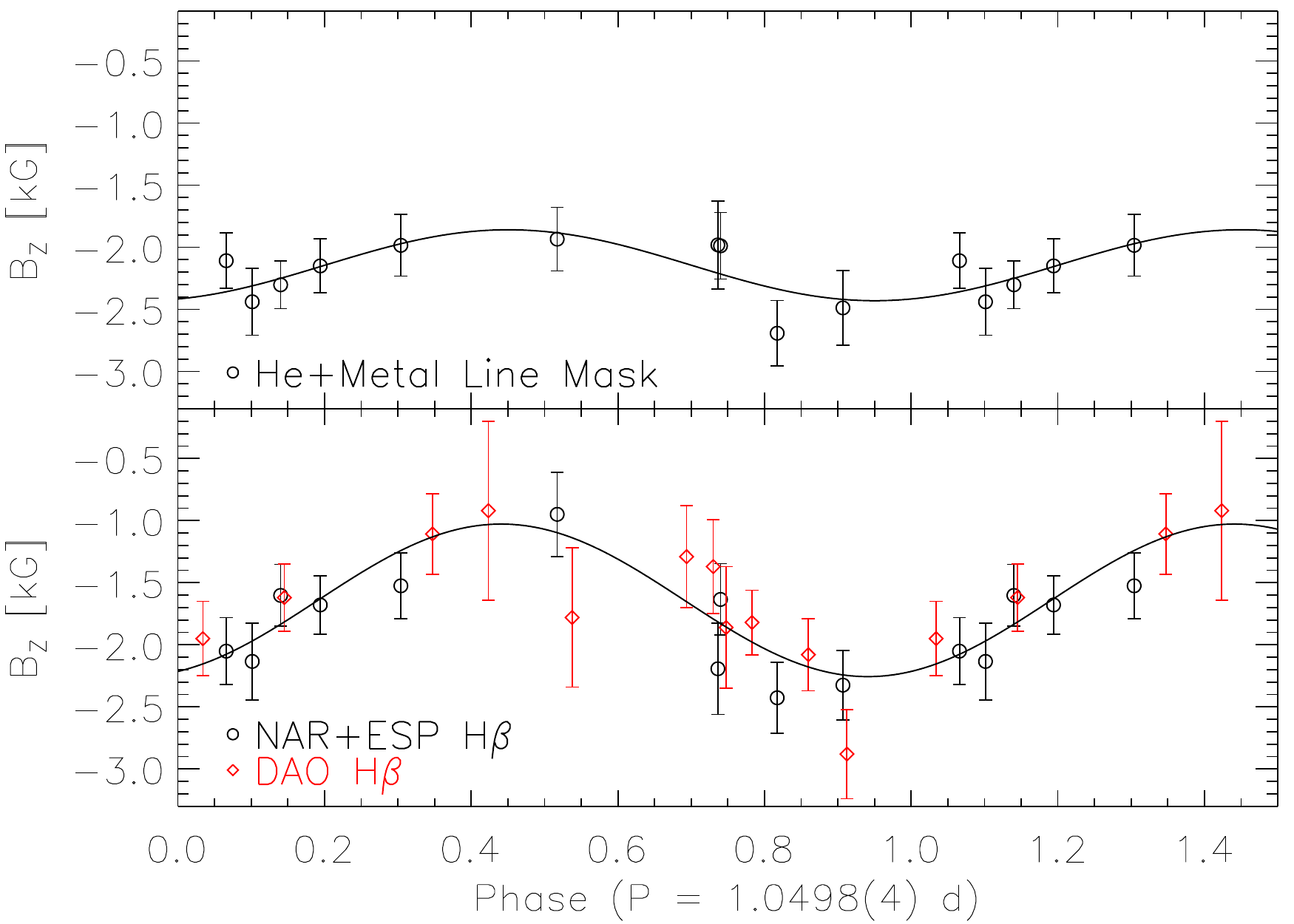}
	\caption{Longitudinal magnetic field measurements phased according to Eqn. 
	\ref{eqn:eph}. The curves correspond to the best-fitting sinusoids. 
	\emph{Top: }$\langle B_z\rangle$ values derived using He+metal LSD line profiles. 
	\emph{Bottom: } $\langle B_z\rangle$ values derived from H$\beta$ Stokes $V$ 
	observations. Black circles correspond to measurements obtained by Narval and ESPaDOnS 
	while red diamonds are those obtained with dimaPol.}
	\label{Bl_phase}
\end{figure}

The longitudinal magnetic field, $\langle B_z\rangle$, is measured through Zeeman 
signatures in the circularly polarized Stokes $V$ spectrum. In order to maximize the 
SNR in these observations, the Least Squares Deconvolution (LSD) cross-correlation 
procedure \citep{Donati1997,Kochukhov2010} was applied to all ESPaDOnS and Narval 
spectra. The LSD line masks are constructed using spectral line lists associated with a 
specified surface gravity, effective temperature, detection threshold (i.e. the central 
depth of the weakest lines included in the mask), and microturbulence ($v_{\rm mic}$) 
obtained from the Vienna Atomic Line Database (VALD2) \citep{Piskunov1995}. In the case 
of HD$\,23478$, a detection threshold of 0.01 and $v_{\rm mic}=0$ were used. Terrestrial 
atmospheric absorption appears as deep, narrow telluric lines in the Stokes $I$ LSD 
profiles that can significantly affect the value of $\langle B_z\rangle$; therefore, 
any lines near spectral regions where tellurics were present were removed.

$\langle B_z\rangle$ measurements are known to vary to some extent depending on which 
element's atomic absorption lines are used to analyze the Zeeman signatures 
\citep[e.g.][]{Pyper1969,Bohlender1987,Yakunin2015}. Several LSD line masks were 
therefore created in order to better evaluate the strength and variability of the 
inferred field -- one which contained only He lines, one in which both H and He lines 
were removed leaving only absorption due to metals, and one containing both He and 
metal lines. 

For each line mask, we adopted normalization values of the wavelength, Land\'{e} 
factor, and line depth of $500\,\text{nm}$, 1.2, and 0.2, respectively. Both the pure He 
and pure metal line masks yielded results similar to the combined He+metal line mask. 
This latter mask resulted in the smallest uncertainties. Examples of the Stokes $V$ and 
$I$, and of the $N$ (circularly polarized intensity, total intensity, and null 
diagnostic) LSD profiles obtained using the He+metal line mask at 3 phases are shown in 
Fig. \ref{LSD}. The profiles corresponding to phases of $0.140$ and $0.817$ were 
obtained by combining two consecutive measurements resulting in a moderately increased 
SNR.

The three LSD profiles shown in Fig. \ref{LSD} demonstrate that no strong variability is 
observed in the He and metal line Zeeman signatures. We also computed the longitudinal 
magnetic field from each LSD profile using Eqn. (1) of \citet{Wade2000} where an 
integration range of $v\in(-140,170)\,\text{km s}^{-1}$ was used. The values 
of $\langle B_z\rangle$ derived from the He+metal mask are reported in Table 
\ref{obs_tbl}. We phased these measurements according to the photometric period of Eqn. 
(\ref{eqn:eph}). The phased longitudinal field measurements calculated using the 
He+metal ($\langle B_z\rangle_{\rm He+metal}$) LSD line mask are shown in Fig. 
\ref{Bl_phase}. 

Measurements of the longitudinal field were also determined using the core of 
the H$\beta$ line. As with the measurements obtained from LSD profiles, the 
first moment method was applied. We integrated the ESPaDOnS and Narval Stokes 
$V$ and $I$ profiles across the full core, from $-200$ to 
$220\,\text{km s}^{-1}$. Stokes $I$ was normalized to the spectral flux 
computed from the 10 pixels located immediately outside the integration range. The 
process used to calculate the DAO longitudinal field measurements is outlined in 
\citet{Monin2012}. The ESPaDOnS and Narval measurements are listed in Table \ref{obs_tbl} 
and the DAO measurements are listed in Table \ref{DAO_obs}.

The phased H$\beta$ longitudinal measurements 
($\langle B_z\rangle_{\rm H\beta}$) show a somewhat larger amplitude variation 
of $0.6\pm0.1\,\text{kG}$ compared with the $0.3\pm0.1\,\text{kG}$ 
amplitude obtained from the He+metal LSD profiles (Fig. \ref{Bl_phase}).

While the sinusoidal fits to the two $\langle B_z\rangle$ data sets appear to 
reproduce the phased measurements reasonably well, only the variability observed in 
$\langle B_z\rangle_{\rm H\beta}$ is statistically significant. Furthermore, 
the phases of the extrema of the sinusoidal fits ($0.4-0.5$ and 
$0.9-1.0$ with uncertainties of about $0.05$) are in agreement amongst 
the two curves. This suggests that the marginal detections of 
$\langle B_z\rangle_{\rm He+metal}$ variations are in fact real.

\begin{table}
	\caption{Calculated values of various parameters associated with the 
	magnetosphere of HD$\,23478$ as derived from He+metal LSD profiles and 
	H$\beta$. For those values dependent on $B_d$, which diverges as 
	$\beta\rightarrow0^\circ$ and $i\rightarrow90^\circ$, only lower limits are 
	provided.}
	\label{mag_tbl}
	\begin{center}
	\begin{tabular*}{0.45\textwidth}{l @{\extracolsep{\fill}} c c}
		\hline
		\hline
		& He+metals$_{\rm LSD}$ & H$\beta$\\
		\hline
		\noalign{\vskip1mm}
		$i\,[\text{deg}]$ 									 	   & \multicolumn{2}{c}{$69^{+21}_{-10}$} \vspace{0.8mm}\\
		$\dot{M}_{B=0}\,[\times10^{-10}\,M_\odot\,\text{yr}^{-1}]$ & \multicolumn{2}{c}{$2.0_{-0.4}^{+0.8}$} \vspace{0.8mm}\\
		$V_\infty\,[\times10^3\,\text{km s}^{-1}]$ 				   & \multicolumn{2}{c}{$1.2^{+0.4}_{-0.3}$} \vspace{0.8mm}\\
		$R_K\,[	R_\ast]$ 									 	   & \multicolumn{2}{c}{$2.9_{-1.2}^{+1.7}$} \vspace{0.8mm}\\
		$\beta\,[\text{deg}]$ 								 	   & $3^{+4}_{-3}$ 	      & $8\pm8$ \vspace{0.8mm}\\				
		$B_d\,[\text{kG}]$ 									 	   & $\geq12.0$ 		  & $\geq9.5$ \vspace{0.8mm}\\
		$\eta_\ast$ 							 	               & $\geq2.1\times10^5$  & $\geq1.3\times10^5$ \vspace{0.8mm}\\
		$R_A\,[R_\ast]$ 									 	   & $\geq21.3$ 		  & $\geq19.0$ \vspace{0.8mm}\\
		$R_A/R_K$ 											 	   & $\geq4.6$ 		      & $\geq4.1$ \vspace{0.8mm}\\
		\hline
	\end{tabular*}\par
	\end{center}
\end{table}

In the context of the Oblique Rotator Model (ORM), periodic variations in 
$\langle B_z\rangle$ can be understood to be due to the rotation of a star having a 
dipolar magnetic field component with an axis of symmetry that is inclined from the 
rotational axis by an angle $\beta$ \citep{Stibbs1950}. Assuming that HD$\,23478$'s 
magnetic field is characterized by an important dipole component, as typically observed 
in hot stars, the faint variability observed in the $\langle B_z\rangle$ measurements 
shown in Fig. \ref{Bl_phase} implies an obliquity angle of $\beta\approx0^\circ$. Other 
explanations for the weak longitudinal field variability can be excluded based on the 
inferred rotational period and observed spectral properties. For instance, regardless of 
the value of $\beta$, no variations in $\langle B_z\rangle$ would be apparent if 
$i=0^\circ$; however, this is inconsistent with the $i=69^{+21\,\circ}_{-10}$ calculated 
in Section \ref{P_rot}. Weak variability would also be observed if the star's rotational 
period is on the order of the observing timescale of $\tau_{obs}\gtrsim30\,\text{d}$. 
This is inconsistent with HD$\,23478$'s inferred $1.0498\,\text{d}$ rotational period 
and relatively large $v\sin{i}$.

In the ORM, the obliquity angle is derived from $i$ along with the best sinusoidal fit 
to the $\langle B_z\rangle$ measurements, given by 
$\langle B_z\rangle=B_0+B_1\sin{2\pi(\phi-\phi_0)}$, using the relation given in Eqn. 3 
of \citet{Preston1967}. A least-squares analysis yielded the fit to 
$\langle B_z\rangle_{\rm He+metal}$ shown in Fig. \ref{Bl_phase}, described by the 
parameters $B_0=-2.1\pm0.1\,\text{kG}$, $B_1=0.3\pm0.1\,\text{kG}$, and 
$\phi_0=-0.3\pm0.4$ where the uncertainties correspond to $1\sigma$ confidence. 
Applying Eqn. 3 of \citet{Preston1967} with the inclination angle derived above and 
$r=0.77\pm0.12$ \citep[Eqn. 2 of ][]{Preston1967} results in an 
obliquity of $\beta=3^{+4\,\circ}_{-3}$. A similar value of 
$8\pm8^{\,\circ}$ was found for $\langle B_z\rangle_{\rm H\beta}$.

With the inclination and obliquity angles known, the polar strength of the surface 
dipole component of the magnetic field can be estimated using the method of 
\citet{Preston1967}. For this calculation, we use a maximum longitudinal field magnitude 
of $|\langle B_z\rangle^{\rm max}|=2.4\pm0.2\,\text{kG}$ as given by the sinusoidal 
fit to $\langle B_z\rangle_{\rm He+metal}$ and a limb-darkening coefficient 
determined by linearly interpolating the values associated with the TLUSTY model 
atmosphere grid yielding a value of $u=0.398$ for the $T_{\rm eff}=20\,\text{kK}$, 
$\log{g}=4.2$ model \citep{Daszynska-Daszkiewicz2011_oth}. This yields a dipole field 
strength of $B_d\geq12.0\,\text{kG}$ where no upper limit is provided due to the 
divergence of the dipole field as $\beta\rightarrow0^\circ$ and $i\rightarrow90^\circ$. 
A relatively small difference in the lower limit of $B_d$ was found 
($9.5\,\text{kG}$) using the combined Narval, ESPaDOnS, and DAO 
$\langle B_z\rangle_{\rm H\beta}$ values. The results of the magnetic analysis are 
summarized in Table \ref{mag_tbl}.

We note that an upper limit on the dipole field strength may be estimated based on the 
fact that Zeeman splitting is not observed in the spectral lines themselves. However, 
given the high $v\sin{i}$ of the star, those constraints are not likely to be very 
helpful. Moreover, the field strength may be better constrained by generating synthetic 
Stokes $V$ profiles \citep[e.g.][]{Petit2012}. On the other hand, its utility would 
likely be hampered by the weak variability of the Stokes $V$ profiles. Such an analysis 
is beyond the scope of this work but could be carried out in the future using a more 
extensive and higher SNR spectropolarimetric data set.

\section{Emission and Variability}
\label{variability}

\begin{figure}
	\centering
	\includegraphics[width=0.99\columnwidth]{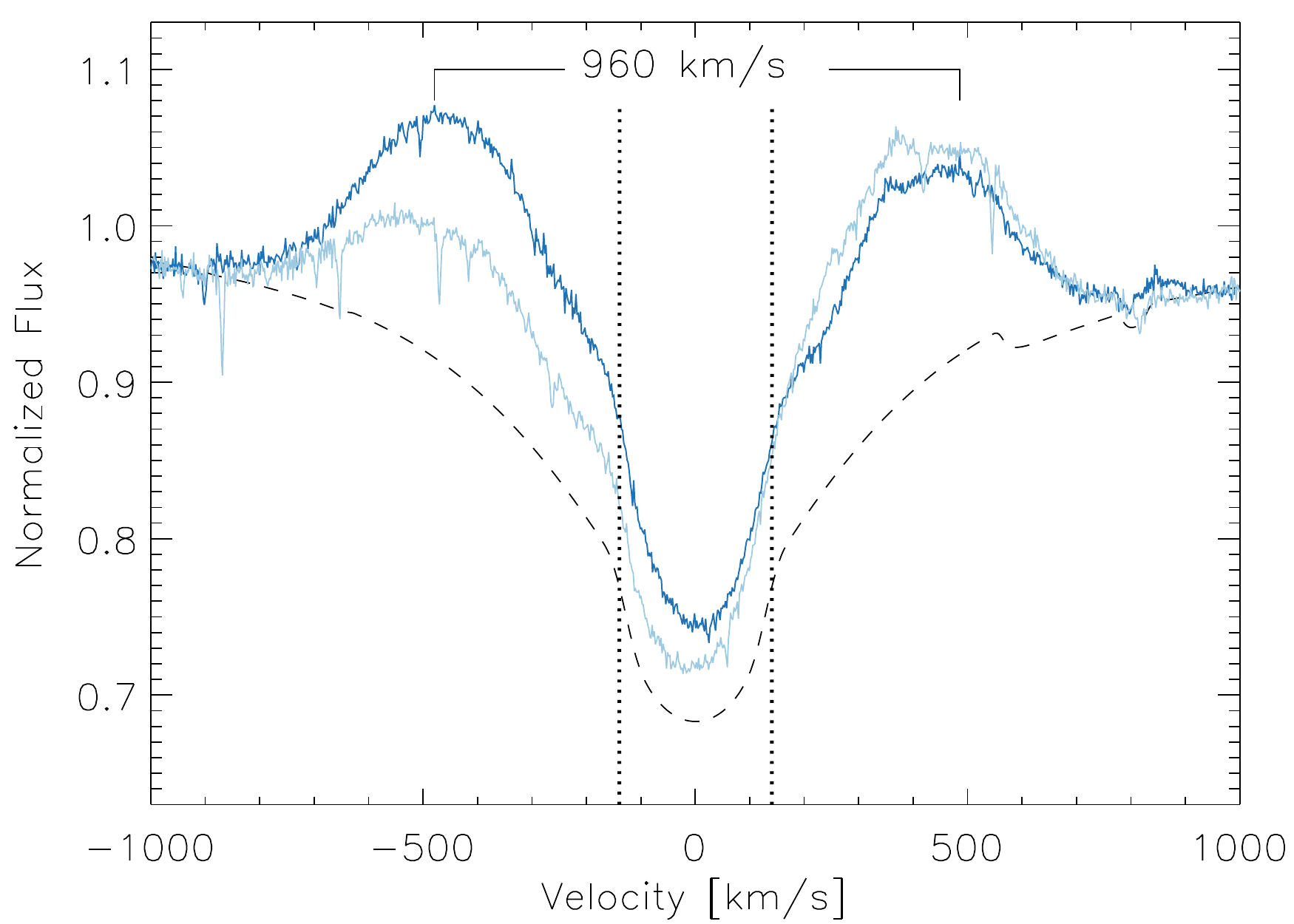}
	\caption{Maximum H$\alpha$ emission (solid dark blue), as observed at a phase of 
	$0.740$, forms a double-horned peak separated by a velocity of 
	$\pm480\,\text{km s}^{-1}$ well beyond the stellar surface indicated by the 
	vertical dotted black lines at $v\sin{i}=\pm140\,\text{km s}^{-1}$. Minimum 
	H$\alpha$ (solid light blue) observed at a phase of $0.516$. The dashed black 
	curve is the $T_{\rm eff}=22\,\text{kK}$, $\log{g}=4.4$ TLUSTY model spectrum 
	discussed in Section \ref{line_fit}.}
	\label{Halpha}
\end{figure}

HD$\,23478$ was originally identified as a potential CM star based on the large 
degree of emission occuring at all near-infrared H lines \citep{Eikenberry2014}. As 
expected, the most significant emission detected in the visible observations presented 
here occurs at the H$\alpha$ line. Fig. \ref{Halpha} shows the minimum and maximum 
emission observed at phases of $0.740$ and $0.516$ and visible at all phases forming a 
double peak with a separation of approximately $\pm480\,\text{km s}^{-1}$. The large 
Doppler shifted velocities of the peaks are more than 3 times the star's projected 
rotational velocity of $v\sin{i}=140\pm10\,\text{km s}^{-1}$ (indicated by the dotted 
vertical lines in Fig. \ref{Halpha}). If we assume that this emission is produced by gas 
bound in corotation with the star, this implies the presence of an emitting plasma at 
$R\gtrsim3R_\ast$.

\begin{figure}
	\centering
	\includegraphics[width=0.99\columnwidth]{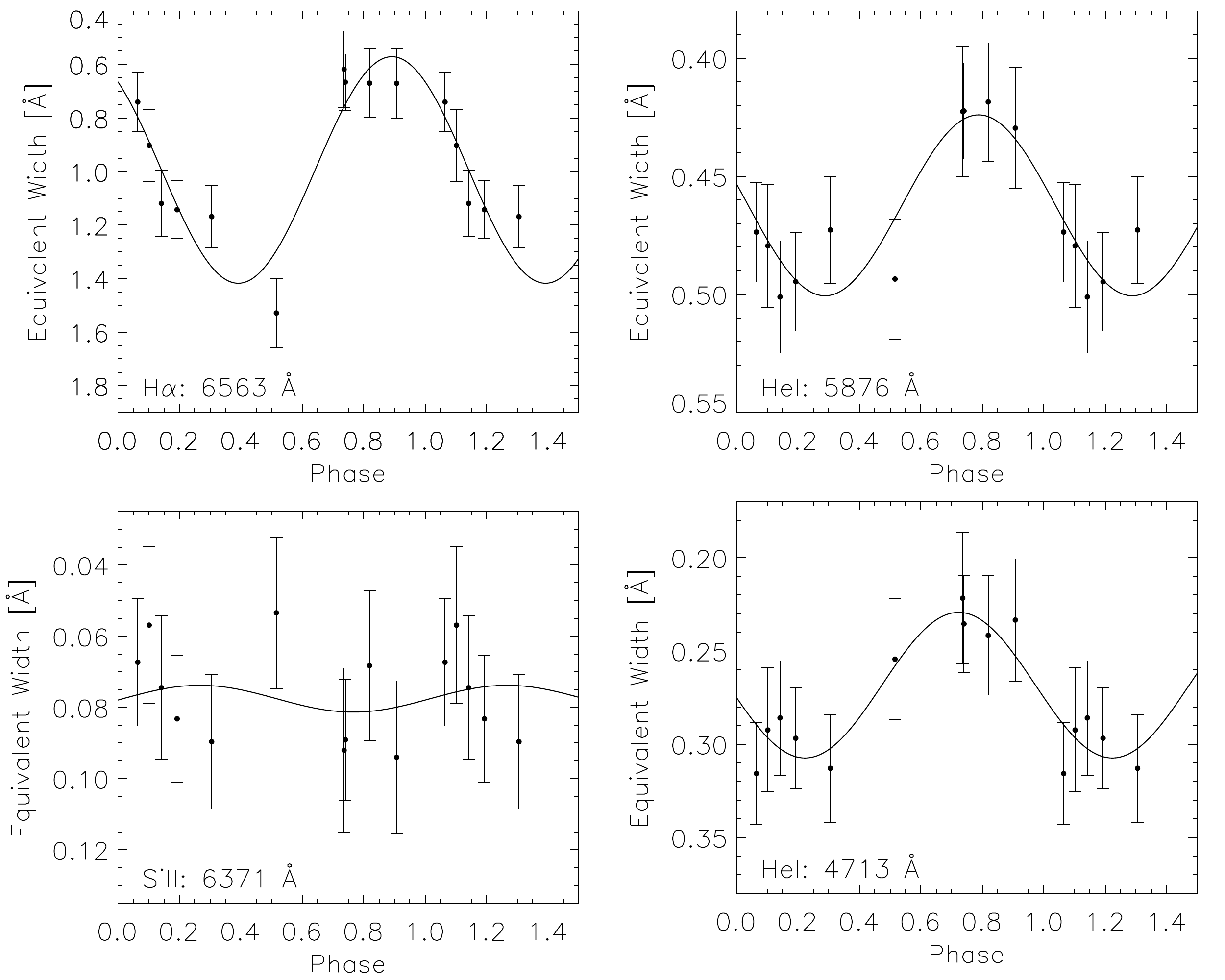}
	\caption{Equivalent widths of variable absorption lines phased with the star's 
	rotational period of $P_{\rm rot}=1.0498\,\text{d}$.}
	\label{EW_comp}
\end{figure}

\begin{figure}
	\centering
	\includegraphics[width=0.99\columnwidth]{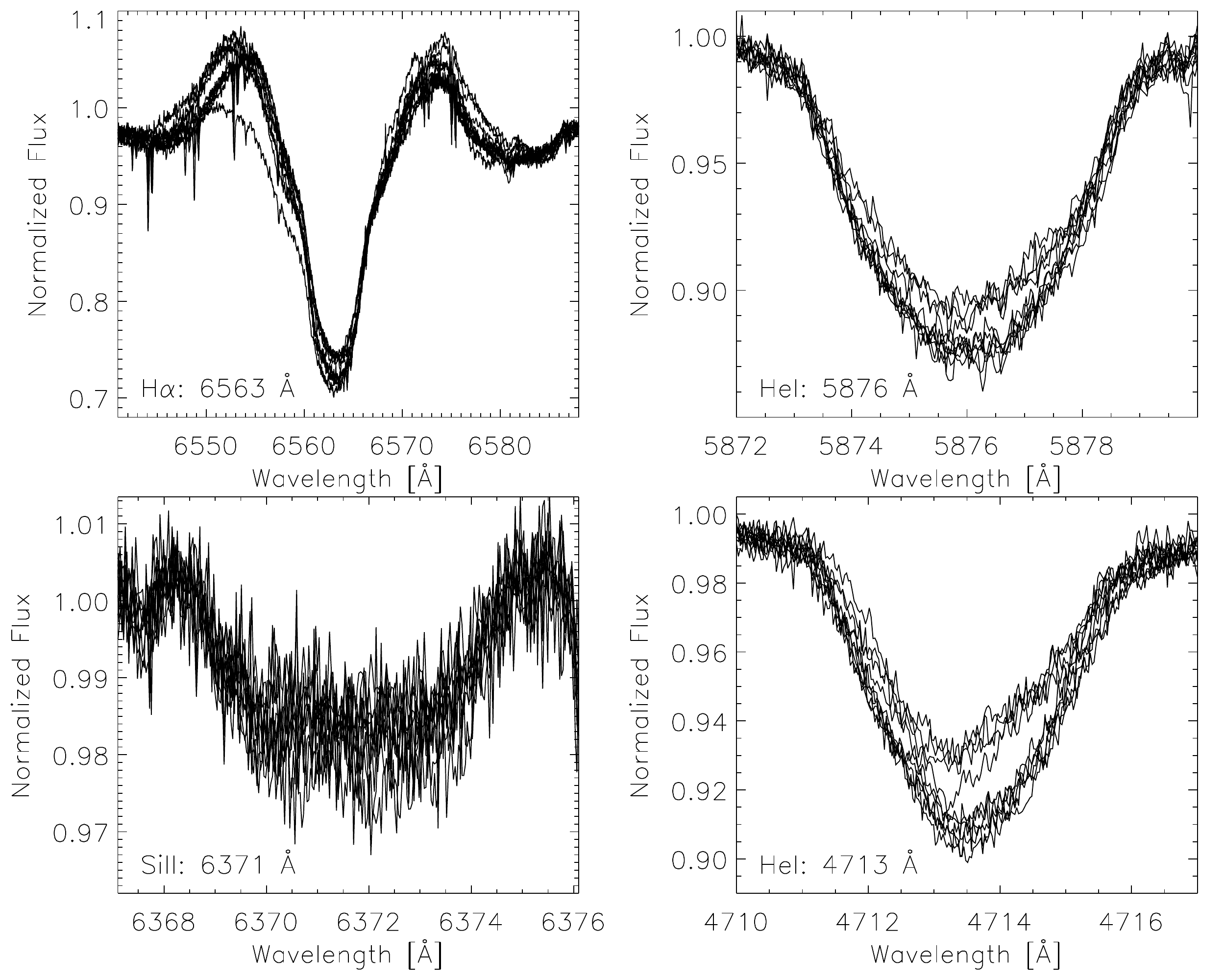}
	\caption{Overlayed line profiles associated with the EWs shown in Fig. \ref{EW_comp} 
	from each of the 12 measurements. Significant variability was found in all identified 
	He~{\sc i} lines.}
	\label{var_comp}
\end{figure}

Although H$\alpha$ presents the largest emission throughout the observed spectra, it 
varies relatively weakly (in comparison to $\sigma$ Ori E, for example) over the 
$1.0498\,\text{d}$ photometric period. This is demonstrated by the equivalent width (EW) 
measurements shown in Fig. \ref{EW_comp} and the overplotted line profiles shown in 
Fig. \ref{var_comp}. Similar to the ORM interpretation of a sinusoidally varying 
longitudinal magnetic field, the continuous and strong emission of CM stars is also 
expected to be rotationally modulated \citep[e.g.][]{Oksala2012}. We find that 
the H$\alpha$ EWs vary coherently when phased by the $1.0498\,\text{d}$ period and is 
therefore consistent with this interpretation.

The clearest example of HD$\,23478$'s spectral variability is found in He line 
EW measurements as demonstrated by He~{\sc i} $\lambda4713$ and He~{\sc i} $\lambda5876$ 
shown in Fig. \ref{EW_comp}. While low SNRs and sparse phase coverage prevent an 
independent determination of $P_{\rm rot}$ from these EW variations, our observations 
are in agreement with a $1.0498\,\text{d}$ period. Furthermore, H$\alpha$ 
and the He lines do not vary coherently when phased by the other candidate 
$1.02454\,\text{d}$ rotational period discussed in Section \ref{P_rot}. This provides 
further evidence in support of a $1.0498\,\text{d}$ rotational period.

The maximum and minimum H$\alpha$ EWs occur at phases $0.39\pm0.05$ and $0.89\pm0.05$, 
respectively. These values are approximately in phase with both the Hipparcos 
photometry and the $\langle B_z\rangle$ measurements. The He~{\sc i} $\lambda5876$ and 
He~{\sc i} $\lambda4713$ EW measurements are in mutual agreement and have maximum and 
minimum values at $0.76\pm0.06$ and $0.26\pm0.06$. They are shifted in phase relative to 
$H_p$ and $\langle B_z\rangle$ by $0.24\pm0.08$, i.e. one-quarter of one cycle.

\section{Magnetosphere}
\label{mag}

The defining characteristic of CM stars aside from the often present strong and broad 
emission lines is their large ratio of the Alfv\'{e}n radius to the Kepler co-rotation 
radius \citep{Petit2013}. The Kepler radius defines the point at which a rigidly 
rotating star's gravitational force is balanced by the outward centrifugal force. Based 
on HD$\,23478$'s inferred mass and rotational period, we calculate a Kepler radius of 
$R_K=2.9_{-1.2}^{+1.7}\,R_\odot$. The Alfv\'{e}n radius ($R_A$) can be estimated from 
the wind confinement parameter ($\eta_\ast$) which depends on the equatorial magnetic 
field magnitude ($B_{\rm eq}$), the stellar radius, mass-loss rate in the absence of a 
magnetic field ($\dot{M}_{B=0}$), and terminal wind speed ($V_\infty$) as derived by 
\citet{ud-Doula2008_oth}. Following the treatment carried out by \citet{Petit2013}, we 
calculate $\eta_\ast$ using the calibration for $\dot{M}_{B=0}$ derived by 
\citet{Vink2000} with $V_\infty=1.3V_{\rm esc}$ for a $T_{\rm eff}=20\,\text{kK}$ star.

Assuming the $B_d$ values calculated in Section \ref{mag_field}, we derived $\eta_\ast$ 
and $R_A$ based on the He+metal and H$\beta$ analyses and find the results to be 
consistent. The larger uncertainties on the H$\beta$ measurements yielded the lowest 
minimum $\eta_\ast$ and $R_A$ at $1.3\times10^5$ and $19.0\,R_\ast$, respectively. With 
a Kepler radius of $R_K=2.9_{-1.2}^{+1.7}\,R_\odot$, we obtain a ratio of 
$R_A/R_K=13.3$ (with a lower limit of 4.6) thereby placing HD$\,23478$ well within the 
CM star regime (i.e. those with $R_A/R_K>1$). The full range of magnetospheric parameters 
calculated from $\langle B_{\rm He+metal}\rangle$ and $\langle B_{\rm H\beta}\rangle$ are 
listed in Table \ref{mag_tbl}.

In the context of all known CM stars, the estimated Kepler and Alfv\'{e}n radii place 
HD$\,23478$ within a region of the magnetic confinement diagram -- Fig. 3 of 
\citet{Petit2013} -- populated by approximately 12 mid to early B-type stars. Six of 
these are reported to exhibit H$\alpha$ emission and only one from this subset -- 
HD$\,142990$ -- has a rotational period shorter than that of HD$\,23478$ 
\citep{Bychkov2005}. The lowest reported dipolar magnetic field strengths of these 12 
CM stars is HD$\,176582$'s $B_p\geq7.0\,\text{kG}$, a Bp star that exhibits 
comparatively weaker H$\alpha$ emission \citep{Bohlender2011}.

We note that HD$\,23478$ exhibits similar physical and spectral properties to the 
well-studied B2Vp star $\sigma$ Ori E. $\sigma$ Ori E has a comparable effective 
temperature of $23\pm1\,\text{kK}$ \citep{Groote1982} and its rotational period and 
projected rotational velocity differ from HD$\,23478$ by as little as 13\% and 6\%, 
respectively \citep{Townsend2010_oth}. H$\alpha$ EW measurements of $\sigma$ Ori E show 
stronger variability $\sim3\,\text{\AA}$ \citep{Oksala2012} compared with the 
$\Delta\text{EW}_{\rm H\alpha}\sim1\,\text{\AA}$ derived for HD$\,23478$. In terms of 
the empirical $R_A/R_K\gtrsim10$ limit noted by \citet{Shultz2014} for the occurence of 
H$\alpha$ emission in CM-hosting stars, these differences in emission properties are 
consistent with $\sigma$ Ori E's likely higher $R_K/R_A$ of approximately $15$ 
\citep{Petit2013}.

Comparisons with the rigidly rotating magnetosphere (RRM) model\footnote{
for a visualization, see: \newline
\href{http://www.astro.wisc.edu/~townsend/static.php?ref=rrm-movies}{www.astro.wisc.edu/\texttildelow townsend/static.php?ref=rrm-movies}}
\citep{Townsend2005} 
suggest that the photometric variability of HD$\,23478$ may be caused by variable 
occultation of the stellar disc by the magnetosphere. In this scenario, the plasma forms 
a circumstellar disk in the magnetic equatorial plane. A non-zero obliquity may then 
allow the plasma to periodically eclipse the stellar disk resulting in an observed 
dimming of the star.

The H$\alpha$ emission and unsigned longitudinal field ($|\langle B_z\rangle|$) 
associated with an $i=70^{\,\circ}$ and $\beta=10^{\,\circ}$ RRM model are predicted to 
be in phase; this is consistent with our analysis of HD$\,23478$. This supports our 
interpretation of the emission and its variability as due to magnetically confined 
plasma in a CM. On the other hand, the model also predicts a maximum photometric 
brightness at the phase of maximum $|\langle B_z\rangle|$, whereas we observe roughly 
the opposite. Therefore, the periodic dimming of the star is likely not related to 
magnetospheric occultation.

Photometric variability may be observed if chemical spots are present on the star's 
surface, which serve to redistribute the star's flux into the UV 
\citep[e.g.][]{Peterson1970,Krticka2013,Krticka2015}. The variable He line profiles 
introduced in Section \ref{variability} provide evidence for He spots on the surface of 
the star. However, the He EWs do not vary in phase with $H_p$ but are instead shifted by 
one quarter of a cycle. Thus, He spots are also not able to fully account for the 
observed $H_p$ variations. On the other hand, spots of other elements (in particular Si 
and Fe) may well also contribute. Such an investigation would be a useful element of a 
future, more detailed study of HD$\,23478$.

\section{Conclusions}
\label{conclusions}

Based on our analysis, we draw the following conclusions:
\begin{enumerate}
	\item HD$\,23478$ is a main sequence magnetic He-strong star
	($T_{\rm eff}=20\pm2\,\text{kK}$) exhibiting peculiar line 
	strengths and distorted line profiles of He, Si, and Fe, indicating likely 
	nonuniform surface distributions of these elements;
	\item the $1.0498(4)\,\text{d}$ period found in Hipparcos epoch photometry 
	is most likely the star's rotational period which confirms the earlier findings 
	of \citet{Jerzykiewicz1993};
	\item the presence of strong H$\alpha$ emission is consistent with the discovery by 
	\citet{Eikenberry2014} of emission in H$\alpha$ and nIR H lines. 
	This emission forms a double peaked profile separated by 
	$v\approx960\,\text{km s}^{-1}$ which likely originates from plasma located at a 
	distance of $R>3\,R_\ast$. We observe significant variability in H$\alpha$ and in 
	various He lines. Measured EWs of these lines are coherently phased by 
	$P_{\rm rot}=1.0498\,\text{d}$;
	\item the surface magnetic field is inferred to have an important dipole 
	component, with a minimum polar strength $9.5\,\text{kG}$, nearly aligned 
	with the star's rotation axis.
\end{enumerate}
Our results unambiguously imply the presence of a centrifugally supported, magnetically 
confined plasma around HD$\,23478$, and therefore confirm the hypothesis of 
\citet{Eikenberry2014} that it is a member of the growing class of magnetic B stars 
hosting centrifugal magnetospheres.

Our analysis yielded a relatively significant discrepancy in the determination of the 
star's effective temperature. IUE spectra used in the SED modelling imply 
$T_{\rm eff}=22\,\text{kK}$ while the observed Balmer lines require a lower temperature 
of $18\,\text{kK}$ to provide the best fit. Chemical peculiarities, line profile 
distortions, and relatively weak and scarce metal lines prevent this ambiguity from 
being confidently resolved within the scope of the present study.

The low inferred obliquity angle of $\beta\leq16^{\,\circ}$ paired with the high 
$69^{+21\,\circ}_{-10}$ inclination angle introduce basic uncertainties in modelling 
the star's magnetic field and magnetosphere. Consequently, only lower limits on the 
Alfv\'{e}n to Kepler radius ratio and the magnetic field's dipole component were 
obtained. We find that the derived $R_A/R_K\geq4.1$ and $B_d\geq9.5\,\text{kG}$ place 
HD$\,23478$ within the more extreme subset of stars hosting centrifugal magnetospheres.

Future work is necessary to more precisely constrain the magnetic field topology, 
magnetospheric properties, and chemical peculiarities of this star. Although the class 
of strongly H$\alpha$ emitting stars hosting centrifugal magnetospheres is growing, the 
total number is still relatively low. Therefore, a detailed characterization of all 
known cases is important in order to better understand the interaction between 
magnetic fields and the winds of hot stars.

\section*{Acknowledgments}

GAW acknowledges support in the form of a Discovery Grant from the Natural Science and 
Engineering Research Council (NSERC) of Canada. AuD acknowledges support by NASA 
through Chandra Award number TM4-15001A and DHC acknowledges support from TM4-15001B 
issued by the Chandra X-ray Observatory Center which is operated by the Smithsonian 
Astrophysical Observatory for and behalf of NASA under contract NAS8-03060. RHDT 
acknowledges support from NASA award NNX12AC72G.

Some of the data presented in this paper were obtained from the Multimission Archive 
at the Space Telescope Science Institute (MAST). STScI is operated by the Association of 
Universities for Research in Astronomy, Inc., under NASA contract NAS5-26555. Support 
for MAST for non-HST data is provided by the NASA Office of Space Science via grant 
NAG5-7584 and by other grants and contracts.

\bibliography{./HD23478.bib,./HD23478-other.bib}
\bibliographystyle{mn2e}

\end{document}